%% file: acl_latex.tex
\documentclass[11pt]{article}

\usepackage[preprint]{acl}

\usepackage{times}
\usepackage{latexsym}

\usepackage[T1]{fontenc}
\usepackage[utf8]{inputenc}

\usepackage{microtype}

\usepackage{inconsolata}

\usepackage{graphicx}
\usepackage{listings}
\usepackage{xcolor}
\usepackage{amsmath}

\usepackage[normalem]{ulem}
\usepackage[T1]{fontenc}
\usepackage{booktabs}
\usepackage{multirow}
\usepackage{pifont}

\usepackage{amssymb}
\makeatletter

\newcommand{\Rmnum}[1]{\expandafter\@slowromancap\romannumeral #1@}
\makeatother
\usepackage{stfloats} 
\usepackage{enumitem}

\usepackage[table]{xcolor}
\usepackage{makecell}
\usepackage{siunitx}
\definecolor{myblue}{RGB}{233, 241, 249}
\definecolor{mygray}{RGB}{99, 110, 114}
\definecolor{myred}{RGB}{255, 118, 117}
\definecolor{myyellow}{RGB}{255, 234, 167}
\definecolor{mygreen}{RGB}{216, 226, 204}
\definecolor{mypurple}{RGB}{162, 155, 254}
\definecolor{mybrown}{RGB}{215, 190, 154}
\definecolor{myorange}{RGB}{255, 220, 190}

\title{FinPos: A Position-Aware Trading Agent System for Real Financial Markets}

\author{
  Bijia Liu \\
  Dongbei University of Finance and Economics \\
  \texttt{qq1403492787@gmail.com}
  \And
  Ronghao Dang$^{*}$ \\
  Alibaba DAMO Academy \\
  \texttt{dangronghao.drh@alibaba-inc.com}
}

\begin{document}
\maketitle
\begin{abstract}
%The exceptional potential of large language models (LLMs) in handling text information has garnered significant attention in the field of financial trading. However, current trading agents primarily focus on single-step trading tasks and lack awareness of continuous position management. Therefore, we propose a position-aware trading task designed to simulate a more realistic market. To address this task, we develop a trading agent system, FinPos, optimized for position management. FinPos is able to interpret various types of market information from a professional perspective, providing a reliable basis for positioning decisions. To mitigate the substantial market risks arising from position fluctuations, FinPos employs dual decision agents. 
%Furthermore, the continuous nature of position management necessitates our adoption of multi-timescale rewards, which in turn empowers FinPos to effectively balance short-term fluctuations against long-term trends.
%Extensive experiments demonstrate that FinPos surpasses state-of-the-art trading agents in the position-aware trading task, which closely mirrors real market conditions. More importantly, our findings reveal that LLM-centered agent systems exhibit a vast, largely unexplored potential in long-term market decision-making.

The exceptional potential of large language models (LLMs) in handling text information has garnered significant attention in the field of financial trading. However, most existing trading agents operate under intraday, independent unit-based trading tasks, where decisions are made as isolated directional actions, and thus lack awareness of continuous position management.
Therefore, we propose a position-aware trading task designed to simulate a more realistic market. To address this task, we propose FinPos, a position-aware trading agent system designed to explicitly model and manage continuous positions. FinPos enhances position awareness through three key mechanisms: (1) professional-level interpretation of heterogeneous market information; (2) a dual-agent decision structure that separates directional reasoning from risk-aware position adjustment; and (3) multi-timescale reward signals, allowing the agent to internalize position awareness through experiential feedback rather than static instructions alone. Extensive experiments demonstrate that FinPos surpasses state-of-the-art trading agents in the position-aware trading task, which closely mirrors real market conditions. More importantly, our findings reveal that LLM-centered agent systems exhibit a vast, largely unexplored potential in long-term market decision-making.
\end{abstract}

%(1) professional-level interpretation of heterogeneous market information to support reliable positioning decisions; (2) a dual-agent decision structure that separates directional reasoning from risk-aware position adjustment, enabling timely responses to sudden market events; and (3) multi-timescale reward signals that provide feedback on both short-term fluctuations and long-term performance, allowing the agent to internalize position awareness through experiential feedback rather than static instructions alone.

\input{sec/1_intro}
\input{sec/2_related_works}
\input{sec/3_task_definition}
\input{sec/4_Architecture_of_FinPos}

\input{sec/5_Experiments}

\input{sec/6_Conclusion}
\input{sec/7_Limitation}

\bibliography{custom}



\appendix
\input{sec/8_Appendix}

\end{document}

%% file: sec/1_intro.tex
\section{Introduction}

\vspace{-0.6em}
\begin{figure*}[t]
\includegraphics[width=\textwidth]{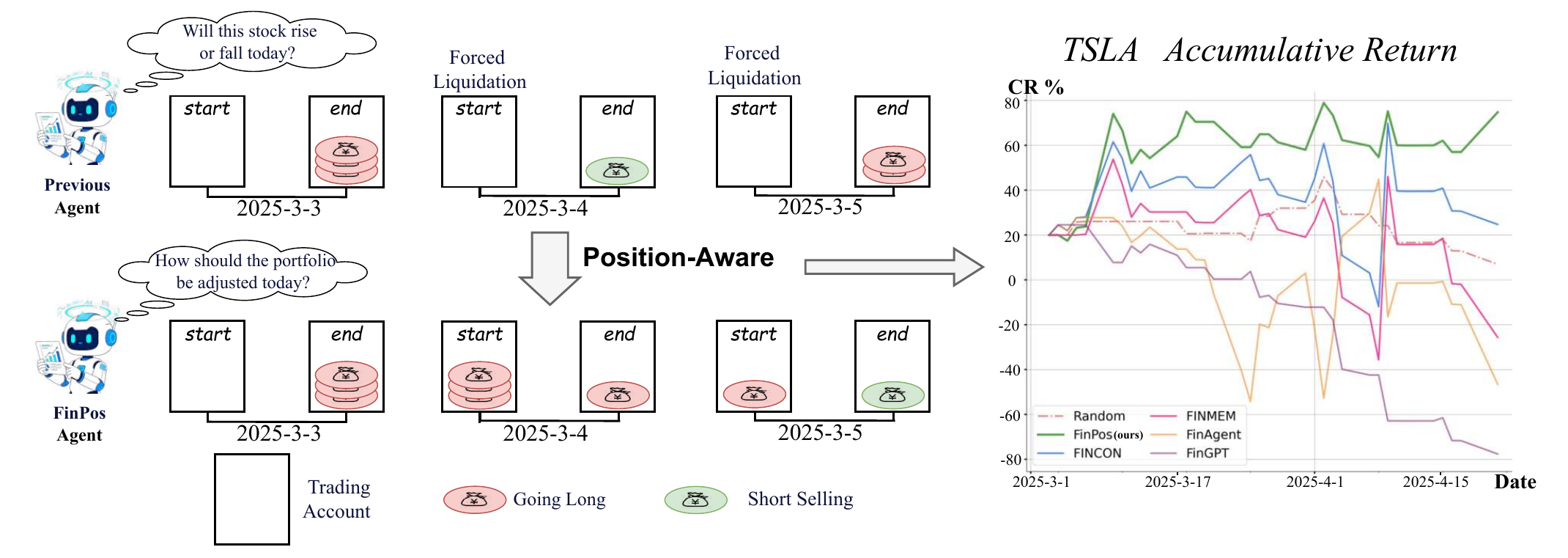}
\caption{With the introduction of position awareness, the agent must not only predict current market trends but also manage the remaining positions in the account. Agents developed for tasks without position awareness are inadequate for addressing the new challenges posed by position-aware trading tasks.} 
\label{intro}
\end{figure*}

%As communication and information technologies evolve, the financial markets are seeing an unprecedented surge in critical, market-moving information~\cite{financial}. Due to inherent cognitive limitations, human traders struggle to process the sheer volume of financial data efficiently~\cite{noise}, which poses a serious challenge for real-time market responses. As a result, automated trading systems~\cite{automated,optimizing,enhancing}, which leverage sophisticated models, are becoming increasingly vital within the financial industry.

An explosion of market information, facilitated by evolving communication technologies, highlights the growing significance of automated trading systems.
In the past decade, deep reinforcement learning (DRL) agents~\cite{reinforcement,RL} gained considerable attention due to their ability to handle large financial datasets. 
However, their inherent difficulty in integrating unstructured text information~\cite{stock} limits their applicability primarily to technical analysis and compromises their interpretability~\cite{survey_DRL}.
% These agents utilize deep neural networks~\cite{DRL} to extract meaningful features, enabling them to approximate complex market dynamics. While DRL agents can learn strategies to maximize investment returns, they still struggle to incorporate unstructured textual information—such as news articles~\cite{news} and financial reports~\cite{financial_reporting}. Furthermore, their limited interpretability~\cite{survey_DRL} makes it difficult for human traders to trust or collaborate with them effectively.

Recently, large language models (LLMs)~\cite{gpt4O,deepseek} have demonstrated remarkable potential in handling text-modal information and solving complex tasks, such as reasoning and decision-making~\cite{llm_decision}. This suggests that LLM-centered agent systems have the potential to transcend the inherent capability boundaries of traditional DRL trading models. Specialized LLM agents designed for financial trading, like TradingAgents~\cite{ta}, FINMEM~\cite{finmem}, and FinAgent~\cite{finagent}, have already achieved groundbreaking results. 
%However, as shown in the left panel of Fig.~\ref{intro}, current LLM agents primarily tackle single-step trading tasks, making them more akin to isolated simulations than realistic financial decision-making systems.
%In real markets, balancing risk and return requires a sophisticated trading agent to shift its focus from short-term market fluctuations to long-term position management. Therefore, we propose a \textbf{position-aware trading task}, which more closely aligns with real-world trading markets than previous setups.
%However, as shown in the left panel of Fig.~\ref{intro}, current LLM-based trading agents are primarily formulated as \textit{Single-Step Trading Tasks}. Their action space is restricted to discrete choices (e.g., \texttt{buy/sell/hold}) with fixed unit sizes, and their internal state does not explicitly track or manage a continuous position. While some works mention "position" in prompts or task descriptions, none implement an explicit position state or a mechanism for dynamic position sizing based on risk constraints. In real financial markets, position management—the continuous adjustment of exposure under risk constraints, evolving market conditions, and long-horizon objectives—is central to sustainable profitability. This fundamental mismatch limits the practical effectiveness of existing agent architectures when deployed in realistic trading scenarios. To bridge this gap, we propose a \textbf{position-aware trading task}, which more closely aligns with real-world trading markets than previous setups.
However, as shown in the left panel of Fig.~\ref{intro}, most existing LLM-based trading agents are formulated under \textit{single-step trading tasks}: each day is treated as an isolated decision episode, with discrete actions and fixed unit sizes. Although some studies mention "position" in prompts, the \textit{single-step trading tasks} fundamentally undermines continuous position management.
%(\texttt{buy}/\texttt{sell}/\texttt{hold})这个不想加了，太占地方了
% this formulation inherently precludes continuous position management—there is no persistent state to optimize over time, no mechanism to adjust exposure dynamically, and no incentive to plan beyond daily profit-and-loss (PnL). 
In contrast, real-world trading is inherently multi-step and stateful: a trader’s current holdings directly shape future risk, opportunity, and decision logic. To bridge this gap, we propose a \textbf{position-aware trading task}, which more closely aligns with real-world trading markets than previous setups.

As shown in the right panel of Fig.~\ref{intro}, existing agent architectures struggle to sustain high yield rates when tasked with position management. This is mainly because they lack three essential capabilities: \textbf{(1) Explicit position representation and exposure control:} Without a continuous position state, existing agents cannot perform risk-aware, dynamic exposure control. \textbf{(2) Long-term planning capability:} Trading tasks that are re-initialized daily are prone to developing agents with a myopic trading perspective. \textbf{(3) In-depth market analysis ability:} Short-term trends are often identifiable through market sentiment, yet a much deeper analysis is essential for understanding long-term growth potential. %To address these challenges, we propose FinPos, a trading agent capable of managing portfolio positions within real financial markets.

To address these challenges, we propose \textbf{FinPos}, a position-aware trading agent designed to manage portfolio positions in realistic financial markets.
FinPos integrates a dual-agent decision architecture with multi-timescale reward mechanism, where a Direction Decision Agent selects discrete actions (\texttt{buy}/\texttt{sell}/\texttt{hold}) and a Quantity and Risk Decision Agent determines transaction volumes.

A defining challenge in \textbf{long-term planning capability} is that the risk and return consequences of an exposure decision are inherently delayed and cumulative—they unfold over multiple trading days rather than within a single step. To explicitly encode this temporal dependency into the agent’s reasoning, FinPos adopts a multi-timescale reward market signal during training, which jointly considers daily immediate profit-and-loss (PnL) and medium- to long-term cumulative returns. This guides the Direction Agent to make non-myopic action choices that balance short-term responsiveness with strategic consistency.

To address \textbf{explicit position representation and exposure control}, the multi-timescale reward framework also feeds long-horizon performance signals into the reflection process of the Quantity and Risk Decision Agent. This enables the agent to internalize the concept of continuous position and understand how its sizing decisions impact portfolio dynamics. Consequently, the Quantity and Risk Decision Agent dynamically determines the specific transaction volume to buy or sell—subject to a CVaR-based upper bound on single-transaction exposure to ensure capital safety.

By observing the reasoning chain of trading agents, we observe that current trading agents often analyze market information superficially and lack causal reasoning abilities. For instance, when trading Tesla stock, the agent may overestimate the relevance of news such as "Musk invests in company A" while dismissing significant news like "major changes in U.S. tariff policy" as irrelevant to trading decisions. This issue arises because the information analysis agent lacks fundamental common sense about the specific forms of information. To enhance \textbf{in-depth market analysis ability}, we build upon the multi-agent information processing paradigm, go further by by injecting domain-specific financial knowledge and causal principles through tailored prompting. This enables FinPos to interpret market signals not just by sentiment, but by fundamental relevance and risk implications.

We conduct extensive experiments using data from multiple real stocks to demonstrate the effectiveness of the FinPos agent in authentic market environments. FinPos outperforms several state-of-the-art (SOTA) financial agents and is the first agent equipped with investment position management capabilities. Moreover, a series of ablation studies underscored the critical importance of risk management, long-term planning, and in-depth market analysis for the agent's ability to manage investment positions effectively.

%% file: sec/2_related_works.tex
\section{Related Works}

\subsection{LLM-Based AI Agents}

% As researchers increasingly recognize the extraordinary capabilities of LLMs in tasks such as summarization~\cite{summarization}, reasoning~\cite{llm_reasoning}, and decision-making~\cite{decision-making}, AI agents built around LLMs are gaining significant attention. 
The evolution of LLMs capabilities has led to a growing interest in LLM-based agents.
Recently, the LLM agents have expanded into various domains, including music~\cite{agentmusic}, healthcare~\cite{medagent}, and research~\cite{researchagent}.
% These LLM agents operate as closed-loop systems, autonomously gathering information, storing memories, reasoning, making decisions, and observing environmental feedback. The advancement of LLMs is propelling AI agents into diverse fields, from open-ended artistic creation to precise medical applications. For instance, MusicAgent~\cite{agentmusic} combines various music-related tools and autonomous workflows to assist artists in music creation. MedAgents~\cite{medagent} use a multi-disciplinary collaboration framework to improve the efficiency of medical diagnosis. Programmer Agent automates the development of software projects. ResearchAgent~\cite{researchagent} helps researchers generate innovative ideas from extensive literature reviews. Personal assistant agents~\cite{personal} engage closely with users, providing solutions for tasks and emotional support. 
These applications illustrate the vast potential of LLM agents in multiple facets of human society. We aim to bring this potential to the financial markets, making LLM agents key players in investment activities.

\subsection{Financial LLM-Based Agents}

The evolution of financial agent architectures has exhibited significant progress through successive innovations and expansions in cognitive modeling and system integration. FINMEM~\cite{finmem} was the first to establish a fundamental trading agent architecture centered around analysis, memory, and decision modules. Building upon this foundation, FinAgent~\cite{finagent} integrate multimodal input to enrich their information sources. FINCON~\cite{fincon} and TradingAgents~\cite{ta} introduced the concept of investment group architectures into agent systems to enhance decision-making reliability. %However
Fundamentally, these agents primarily address a simplified financial game focused on predicting next-day price movement. In real markets, however, agents must not only needs to determine the investment direction, but also decide when to adjust exposure and how much to allocate. Therefore, this paper adopts a position-aware trading task that more closely approximates real market environments.

%% file: sec/3_task_definition.tex
\section{Task Definition}

\subsection{Single-Step Trading Task}

We define a baseline environment as a \textit{Single-Step Trading Task}, where the agent's position is automatically liquidated at each timestep, and no position is carried over. Within this framework, the agent makes a trading decision $a_t \in \{-1, 0, 1\}$ at each timestep $t$, corresponding to selling, holding, or buying, respectively. Each decision assumes immediate liquidation at the next timestep, thereby making each action independent, with no continuation of position states. The cumulative return $R$ is the sum of the logarithmic returns at each step:

\vspace{-0.7em}
\begin{equation}
R = \sum_{t = 1}^{N} r_t, \quad 
r_{t} = a_{t} \times \log 
\frac{\text{price}[t+1]}{\text{price}[t]}
\end{equation}

This setup is widely adopted in studies such as FINMEM~\cite{finmem} due to its simplification of trading processes, which stabilizes the training environment and allows more straightforward model evaluation. However, this simplification comes with significant limitations: it completely ignores the continuity of holding positions and the gains and losses arising from holding positions in real trading, thus cannot capture the associated compound returns or the risk exposure carried across periods. Without modeling positions, the agent has no basis for risk control or for detecting market trends, both of which are central to real trading.
% These omissions weaken the agent’s learning and generalization in essential skills such as position management and risk control.
More critically, by reducing trading to short-term, step-by-step predictions, such a setup underutilizes the natural strengths of LLMs in reasoning over longer horizons. Consequently, advancing financial trading agents towards practical application requires a task more closely mirrors real market conditions.

\subsection{Position-Aware Trading Task}
\label{sec:task}

Compared to the aforementioned simplified setup, this paper proposes a trading task configuration that aligns more closely with real market behavior, \textit{Position-Aware Trading Task}. In this environment, the agent makes a trading decision on each trading day (each timestep \( t \) based on the closing price) and explicitly maintains its position state. The agent's decisions no longer involve automatic liquidation; rather, it must consider the continuation and adjustment of current positions in future timesteps. Returns $R$ are calculated from the cumulative logarithmic returns based on position:

\vspace{-0.7em}
\begin{equation}
R = \sum_{t = 1}^{N} r_t, \quad 
r_{t} = \text{pos}_{t} \times \log \frac{\text{price}[t+1]}{\text{price}[t]}
\label{cr}
\end{equation}

Where \(\text{pos}_{t}\) represents the agent's current position state, dynamically updated based on past trading decisions. This setup not only more accurately models the continuity and trend dynamics of holding positions in trading but also allows the agent to naturally incorporate key capabilities in strategy learning, such as position control, risk exposure management, and entry-exit timing. 
The \textit{Position-Aware Trading Task} provides the agent with a more challenging and practically meaningful learning objective, offering stronger potential for strategy generalization and real-world applicability.

Additionally, this task better aligns with the capability characteristics of trading LLM agents. Compared to traditional quantitative models, LLMs excel at extracting long-term trends and underlying causal structures from complex semantic information rather than performing high-frequency precise numerical optimization. This capability gives them greater potential in mid-to-long-term strategy formulation. Under the \textit{Single-Step Trading Task} setup, the agent primarily focuses on the price direction of the next step, which somewhat forces the LLM struggling to build coherent strategies or cognition. By introducing position continuity and strategy coherence, the \textit{Position-Aware Trading Task} provides LLM agents with a more fitting testing platform, fully unlocking their potential in long-term decision-making and complex information integration. Furthermore, this setup enables agents to construct an internally consistent position management logic centered around fundamentals, market sentiment, and macro semantic information. The combination of long-term and semantic-driven strategies is precisely where LLMs hold their advantages in financial trading tasks.

%% file: sec/4_Architecture_of_FinPos.tex
\section{Architecture of FinPos} 

%The architecture of FinPos (Fig.~\ref{fig:archi}) consists of three core modules: the market signal processing and analysis module, the trading decision module, and the multi-timescale reward reflection module. The signal processing module introduces a two-level agent hierarchy—filtering and analysis—to process raw data, with the results stored in a memory layer. The decision module advances beyond prior research that only predicts direction by introducing a position-aware mechanism. A direction decision agent determines the trading orientation, while a quantity decision agent dynamically adjusts position size in response to market risk exposure. The reward module adopts a multi-timescale reward feedback mechanism to enhance the long-term behavioral stability of the model and its responsiveness to trading opportunities. The details of these three modules are presented in the subsequent sections, while all agent prompts are provided in the Appendix ~\ref{appendix:agent_prompts}.

\begin{figure*}[t]
\includegraphics[width=\textwidth]{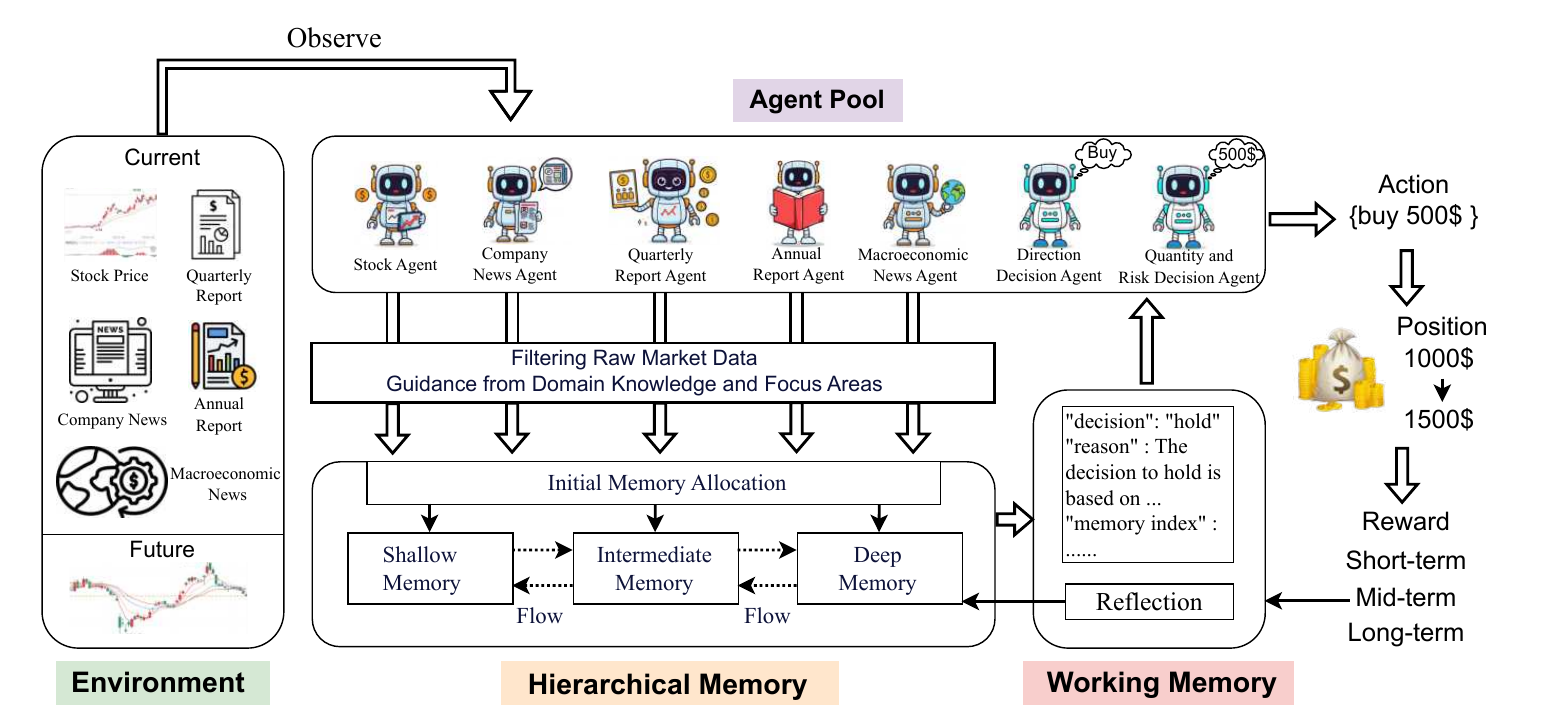}
\caption{\textbf{Architectural Details of FinPos:} Initially, multiple analysis agents leverage domain knowledge to gather diverse information from the environment, subsequently storing it in the memory module. The memory module utilizes a memory allocator to distribute the acquired information across memory layers of varying depths. Subsequently, the most pertinent information for the current decision is placed into working memory, where dual decision agents generate trading actions, while multi-timescale reward signals support reflective updates that consolidate experiential knowledge into deeper memory layers.} 
\label{fig:archi}
\end{figure*}

The architecture of FinPos (Fig.~\ref{fig:archi}) consists of three core modules. The \textbf{market signal processing module} adopts a two-level agent hierarchy, consisting of filtering agents and analyst agents, to distill raw market data into structured, decision-relevant signals. 
% The \textbf{trading decision module} implements a dual trading decision mechanism to explicitly model position dynamics. A direction decision agent determines the trading orientation, while a quantity decision agent dynamically adjusts position size in response to market risk exposure. 
The \textbf{trading decision module} employs a dual decision framework that explicitly decouples quantity decision-making from directional determination.
The \textbf{multi-timescale reward module} adopts a multi-timescale reward feedback mechanism to experientially internalize the long-term risk and return implications of position changes. The details of these three modules are presented in the subsequent sections, while all agent prompts are provided in the Appendix ~\ref{appendix:agent_prompts}.

\vspace{-0.8em}
%Subsequently, the most pertinent information for the current decision is placed into working memory, where dual decision agents output trading actions. Finally, multi-timescale rewards guide reflection, storing experiential knowledge into deeper memory layers.

\subsection{Market Signal Processing and Analysis}

FinPos draws inspiration from institutional investment workflows. It assigns specialized agents to distinct information domains forming a division of labor similar to that in private equity firms. For each domain, we further establish a two-tier structure: Signal Processing Agents and Analysis Agents.

\textbf{Signal Processing Agents:} These agents are responsible for preprocessing and filtering high-noise, heterogeneous market data streams. Through domain-specific prompts and heuristic rules, the agents clean, compress, and prioritize information by relevance and importance. For example, large volumes of low-value or weakly related news are downweighted or discarded, while impactful macroeconomic policies or firm-level events are highlighted and passed on to subsequent modules.

\textbf{Analysis Agents:} These agents analysis the filtered data.  We observe that large language models often exhibit financial hallucinations in market scenarios, making it difficult for them to accurately capture the key factors influencing stock prices. They tend to rely solely on explicit keywords in news while overlooking the underlying market logic. For example, an LLM may fail to recognize that news about the S\&P 500 is strongly correlated with the performance of leading U.S. stocks. To mitigate this issue, we explicitly inject financial knowledge into the prompts, thereby strengthening their causal awareness. As financial reasoning is gradually integrated into the prompts, the agents evolve from mechanical summarizers into analysts with genuine financial reasoning ability, capable of generating deeper insights and producing more accurate interpretations of market dynamics.

Subsequently, as illustrated in Fig.~\ref{fig:archi}, all analytical results are aggregated into a Hierarchical Memory Module. In this module, important long-term information (e.g., annual reports) is allocated to deep memory, while volatile short-term information (e.g., corporate news) is stored in shallow memory. The hierarchy of memory is not static but dynamically adjusted through post-decision reflection: memories that repeatedly prove their validity are gradually migrated into deeper layers, thereby increasing their weight in future decision-making.

\subsection{Dual Trading Decision}
\label{sec:pos}

In real-world trading, an account’s current holdings directly reflect its risk exposure. Professional traders adjust their buy–sell decisions dynamically based on existing positions to maintain a relative balance between risk and return. FinPos adopts a dual-agent framework that embeds position awareness into both \textbf{the Direction Decision Agent} and \textbf{the Quantity and Risk Decision Agent}. By integrating multi-source information retrieved from the memory layer, the dual-agent framework enables a decision-making mechanism that more closely mirrors practical trading logic.

\subsubsection{The Direction Decision Agent} 
\label{sec:decision}
Incorporating position awareness requires integrating longer-term information. The direction decision agent combines structured memory (news, financial reports, and past actions) with the current portfolio state and explicitly specifies the strategic intent of each action—whether it represents a long-term position-building move or a short-term tactical adjustment to exploit local trends. This explicit articulation not only enhances interpretability for human supervisors but also provides downstream quantity agents with a clear strategic context. To prevent the agent from over-relying on short-term fluctuations, we design prompts that encourage it to factor in longer horizons, thereby gradually cultivating its long-term planning capability.

\subsubsection{Quantity and Risk Decision Agent}

FinPos introduces the Quantity and Risk Decision Agent, which determines the specific trade size after a directional decision has been made. This agent determines trade sizes through an explicit risk-aware mechanism that jointly incorporates the current position state, structured memory, and CVaR-based risk references (Conditional Value at Risk~\cite{CVaR}; see Appendix~\ref{appendix:cvar}), with per-step transaction sizes capped by the 95\% CVaR to control exposure under volatility. This design ensures that every trade size is grounded in the agent’s current position and risk tolerance, making exposure control an explicit, first-class component of FinPos’s decision process.

\subsection{Multi-Timescale Reward Design}
\label{sec:score}

To guide the agent toward non-myopic trading behavior, we design a trend-aware reward function that leverages \textit{multi-timescale market signals}. Specifically, we calculate three future price trends in each timestep $t$.
$M_{t}^{s} = price[t+1] - price[t]$: 1-day trend (short-term),
$M_{t}^{m} = price[t+7] - price[t]$: 7-day trend (mid-term),
$M_{t}^{l} = price[t+30] - price[t]$: 30-day trend (long-term).
We define the multi-timescale score $M_t$ as:

\vspace{-0.5em}
\begin{equation}
    M_t = M_{t}^{s} + M_{t}^{m} + M_{t}^{l}
\end{equation}
This score represents the aggregated expected trend across multiple timescale (e.g., 1-day, 7-day, and 30-day). The reward at time $t$, $Reward_{t}$ is defined as follows:

\vspace{-0.7em}
\begin{equation}
Reward_{t} = \left\{\begin{matrix}
-(M_{t})^{2}, & pos_{t} =pos_{t-1} \\  
pos_{t} \times M_{t}, & otherwise \\
\end{matrix}\right.
\end{equation}

\vspace{-0.7em}
\begin{equation}
    pos_{t} = pos_{t - 1} + d_t \times q_t 
\end{equation}
$d_t$ and $q_t$ denote the direction and quantity of the purchasing decision at the current time step, respectively.
Crucially, this reward is used only during training-time simulation, no future-dependent signals are accessed at test time. 

It serves two complementary roles in guiding the dual-agent architecture:
\textbf{(1) Guiding the Direction Agent:} The term $\texttt{pos}_t \times M_t$ provides positive reinforcement when the agent’s directional decision (\texttt{buy}/\texttt{sell}/\texttt{hold}) aligns with the multi-horizon trend $M_t$. This encourages the Direction Agent to move beyond short-term noise and base decisions on longer-term market signals.
\textbf{(2) Guiding the Quantity Agent:}  During LLM reflection, the Quantity Agent uses this reward to learn when to adjust exposure. Through this process, the agent internalizes the concept of position as a dynamic state. In our experiments, we find that during periods of high volatility, agents tend to maintain a inactive state. To address this, we introduce a quadratic penalty on $|M_t|$. whenever the position remains unchanged. This discourages missed opportunities in volatile markets while also preventing excessive trading in stable periods. 

%The multi-timescale reward is used only during training-time simulation to guide LLM reflection. At test time, no future-dependent signals are accessed.

%This reward design serves two purposes:
%1. \textbf{Encourage trend alignment:} The agent receives positive reinforcement when its predicted direction matches the actual market movement. In this way, the agent is guided away from overfitting to daily fluctuations and instead learns to associate price trajectories with textual information and broader market trends, enabling more robust causal inference in decision-making.  
%2. \textbf{Penalty for passivity:} In our experiments, we find that during periods of high volatility, agents tend to maintain a inactive state. To address this, we introduce a quadratic penalty on $|M_t|$. whenever the position remains unchanged. This discourages missed opportunities in volatile markets while also preventing excessive trading in stable periods.

%% file: sec/5_Experiments.tex
\section{Experiments}

\subsection{Experimental Setup}
\subsubsection{Datasets}
Our research utilizes actual financial data sourced from publicly authoritative providers and encompasses various forms of market information. (1) Stock prices, obtained from Yahoo Finance, include daily open-high-low-close-volume (OHLCV) data. (2) Company news, retrieved using the Finnhub stock API, includes articles' related company name, publication date, headline, and summary text, processed for sentiment and semantic relevance. (3) Macroeconomic news, also retrieved via the Finnhub stock API, contains publication dates, headlines, and summary texts for each article. (4) 10-Q (quarterly reports) and 10-K (annual reports), accessed through the U.S. Securities and Exchange Commission (SEC) EDGAR API. These documents provide financial information that publicly listed companies are required to disclose, and standardized into a daily time series format. 

\subsubsection{Comparative Methods}
To assess the effectiveness of FinPos, we conduct comparative experiments with four LLM-based trading agents: FinGPT~\cite{FinGPT}, FINMEM~\cite{finmem}, FinAgent~\cite{finagent}, and FINCON~\cite{fincon}; three deep reinforcement learning (DRL) methods (A2C, PPO, DQN); two rule-based methods (MACD, RSI); and two market baselines (Random). All compared models are evaluated using identical data splits and executed under the same market conditions. Additional details are provided in Appendix~\ref{appendix:es}.

\subsubsection{Evaluation Metrics}
We evaluate performance using cumulative return (CR\%), sharpe ratio (SR), and maximum drawdown (MDD\%).
CR reflects overall profitability, while SR measures the risk-adjusted return by showing how much excess return it generates per unit of total risk. 
MDD quantifies the worst-case loss over the trading horizon, which in the position-aware trading task 
directly reflects exposure and vulnerability under held positions.  
CR has been introduced earlier (see Eq.~\eqref{cr}), and the formulas for SR and MDD are provided in Appendix~\ref{app:metrics}.

\subsubsection{Implementation Details}
All LLM trading agents are deployed using GPT-4o~\cite{gpt4O} with a temperature setting of 0.7. All models are evaluated under the position-aware trading task formally defined in Sec.~\ref{sec:task}, ensuring profits and risks are calculated with explicit holding dynamics. Training covered Jan 2024–Feb 2025; testing spanned Mar–Sep 2025, covering the U.S. election period and other major macroeconomic events, making it a representative and challenging evaluation setting.

\begin{table*}[t]
\centering
\caption{Performance comparison of different models on five stocks}
\label{tab:model_comparison}
\resizebox{\textwidth}{!}{%
\begin{tabular}{l S S S S S S S S S S S S S S S S}
\Xhline{2\arrayrulewidth}
\multirow{2}{*}{\textbf{Models}} & \multicolumn{3}{c}{\textbf{TSLA}} & \multicolumn{3}{c}{\textbf{AAPL}} & \multicolumn{3}{c}{\textbf{AMZN}} & \multicolumn{3}{c}{\textbf{NFLX}} & \multicolumn{3}{c}{\textbf{COIN}} \\
\cmidrule(lr){2-4} \cmidrule(lr){5-7} \cmidrule(lr){8-10} \cmidrule(lr){11-13} \cmidrule(lr){14-16}
 & CR\%↑ & SR↑ & {MDD\%↓} & CR\%↑ & SR↑ & {MDD\%↓} & CR\%↑ & SR↑ & {MDD\%↓} & CR\%↑ & SR↑ & {MDD\%↓} & CR\%↑ & SR↑ & {MDD\%↓} \\
\Xhline{2\arrayrulewidth}
\rowcolor{myred!10}
\multicolumn{16}{c}{\textit{Market Baseline}} \\ \Xhline{2\arrayrulewidth}
Random    & -25.81 & -0.62 & 62.10 & -16.46 & -0.23 & 39.40 & -7.80 & -0.34 & 61.20 & -3.04 & -0.02 & 32.60 & 1.27 & 0.14 & 57.60 \\
\Xhline{2\arrayrulewidth}
\rowcolor{myyellow!30}
\multicolumn{16}{c}{\textit{Rule-Based Methods}} \\ \Xhline{2\arrayrulewidth}
MACD~\cite{macd}   & -46.25 & -0.55 & 74.20 & -29.61 & -0.42 & 51.40 & -38.14 & -0.47 & 70.30 & -27.45  & -0.23 & 47.92 & -49.87  & -0.54 & 70.83  \\
RSI~\cite{rsi}    & -45.06 & -0.52 & 72.80 & -33.28 & -0.48 & 49.70 & -35.10 & -0.42 & 68.20 & -26.98  & -0.21 & 46.08 & -53.65  & -0.63 & 66.37  \\
\Xhline{2\arrayrulewidth}
\rowcolor{mygreen!50}
\multicolumn{16}{c}{\textit{Reinforcement Learning Methods
}} \\ \Xhline{2\arrayrulewidth}
A3C~\cite{a3c}    & -86.99 & -0.94 & 96.60 & -66.13 & -0.75 & 73.40 & -56.63 & -0.62 & 72.80 & -35.24  & -0.51 & 64.60 & -58.96  & -0.69 & 68.60  \\
DQN~\cite{dqn}    & -71.25 & -0.78 & 80.62 & -47.96 & -0.57 & 56.30 & -43.87 & -0.48 & 68.10 & 0.78  & 0.11 & 41.28 & -19.25  & -0.26 & 48.41  \\
PPO~\cite{ppo}    & -56.72 & -0.48 & 69.88 & -35.50 & -0.42 & 51.40 & -28.45 & -0.39 & 64.90 & -11.92  & -0.26 & 51.30 & 9.32  & 0.34 & 56.39 \\
\Xhline{2\arrayrulewidth}
\rowcolor{myblue}
\multicolumn{16}{c}{\textit{LLM-Based Agent
}} \\ \Xhline{2\arrayrulewidth}
FinGPT~\cite{FinGPT}    & -89.36 & -0.98 & 94.36 & -80.37 & -1.01 & 74.60 & -60.14 & -0.67 & 70.30 & -26.01  & -0.40  & 57.30 & -42.63 & -0.34 & 67.09 \\
FinAgent~\cite{finagent}  & -65.07 & -0.76 & 85.65 & -51.10 & -0.69 & 49.50 & -43.29 & -0.47 & 61.20 & -18.71  & -0.22  & 41.50 & 2.76 & 0.23 & 56.50 \\
FINMEM~\cite{finmem}    & -36.48 & -0.45 & 72.10 & -38.94 & -0.52 & 44.82 & -27.61 & -0.33 & 66.10 & -33.75  & -0.37  & 60.02 & -2.13  & -0.12  & 51.28 \\
FINCON~\cite{fincon}    & 19.67  & 0.36  & 59.13 & -2.81 & -0.06 & 31.03 & 7.96  & 0.12 & 34.77 & 6.12  & 0.17  & 34.70 & 21.35  & 0.63  & 46.10 \\
\Xhline{2\arrayrulewidth}
\rowcolor{mypurple!20}
\textbf{FinPos}    & \textbf{62.15} & \textbf{0.68} & \textbf{42.34} & \textbf{36.31} & \textbf{0.43} & \textbf{27.53} & \textbf{30.35} & \textbf{0.34} & \textbf{18.44} & \textbf{28.65} & \textbf{1.02} & \textbf{20.05} & \textbf{54.36} & \textbf{0.87} & \textbf{34.05} \\
\Xhline{2\arrayrulewidth}
\end{tabular}%
}
\end{table*}

\subsection{Main Results}
We evaluated FinPos on five representative stocks: TSLA, AAPL, AMZN, NFLX, and COIN. As shown in Tab.~\ref{tab:model_comparison}, FinPos consistently outperforms all baseline methods, achieving higher CRs, SRs, and lower MDDs across all assets. On high-volatility stocks such as TSLA, AAPL, and COIN, FinPos exhibits strong robustness under turbulent market conditions. While DRL baselines (A3C, DQN, PPO) and LLM-based agents (FinGPT, FinAgent) suffer substantial losses and severe drawdowns (e.g., A3C loses over 80\% on TSLA with an MDD of 96.6\%), FinPos consistently maintains positive returns, achieving CRs of 62.15\% and 36.31\% on TSLA and AAPL, respectively, with well-controlled downside risk.
For assets with clearer trends, such as AMZN and NFLX, FinPos continues to deliver stable and risk-adjusted performance. FinPos achieves a CR of 30.35\% on AMZN with an MDD of 18.44\%, and a CR of 28.65\% on NFLX with a high SR of 1.02, outperforming all baseline agents in both return and risk control.

\subsection{Ablation Studies}
We select TSLA, AAPL, and AMZN as test assets due to their rich coverage of news articles, earnings events, and macroeconomic sensitivities. We assess the contribution of each FinPos component through ablation experiments. Tab.~\ref{tab:ablation_single} reports results under different module removals, showing that the full model consistently achieves the best performance. Extended ablation results are provided in Appendix~\ref{appendix:Ablation}.

\begin{itemize}[left=0.25cm]
\item \textbf{Multi-Timescale Reward (MTR)} is the cornerstone of our position-aware framework. It provides a self-supervised signal that aligns short-term actions with long-horizon outcomes, enabling the agent to internalize the cumulative impact of position decisions. 
\item \textbf{Quantity and Risk Decision Agent (QRA)} is essential for explicit position management. It transforms the agent from a unit-action trader into a continuous-position controller. Removing QRA forces fixed-unit trades, eliminating dynamic exposure control.
\item \textbf{Market Signal Processing (MSP)} acts as a preprocessing filter that scores raw market data by \textit{relevance} and \textit{importance} before downstream agents analyze them, ensuring that only high-signal information enters the decision pipeline.
\end{itemize}

\vspace{-0.6em}
\begin{table*}[t]
\centering
\footnotesize
\caption{Ablation Results on TSLA, AAPL and AMZN}
\label{tab:ablation_single}
\resizebox{\textwidth}{!}{%
\begin{tabular}{c c c | c c c | c c c | c c c}
\toprule
\textbf{MTR} & \textbf{QRA} & \textbf{MSP} & \multicolumn{3}{c|}{\textbf{TSLA}} & \multicolumn{3}{c|}{\textbf{AAPL}} &\multicolumn{3}{c}{\textbf{AMZN}} \\
           &              &            & CR\%↑ & SR↑ & MDD\%↓ & CR\%↑ & SR↑ & MDD\%↓ & CR\%↑ & SR↑ & MDD\%↓ \\
\midrule
           & \checkmark  & \checkmark   & 18.73  & 0.35 & 57.87    & 12.93 & 0.20 & 35.63 & 9.75 & 0.21 & 26.20    \\
\checkmark          &             & \checkmark           & 53.57     & 0.49 & 62.65      & 29.30     & 0.31 & 39.29   & 27.85 & 0.43 & 30.29   \\
\checkmark          & \checkmark           &             & 58.34     & 0.63 & 45.40      & 34.09     & 0.39 & 29.87  & 28.50 & 0.32 & 19.87    \\
\checkmark          & \checkmark            & \checkmark                      & \textbf{62.15} & \textbf{0.68} & \textbf{42.34} & \textbf{36.31} & \textbf{0.43} & \textbf{27.53} & \textbf{30.35} & \textbf{0.34} & \textbf{18.44}
\\
\bottomrule
\end{tabular}
}
\end{table*}

\begin{figure}[t]
    \centering
    \includegraphics[width=0.95\linewidth]{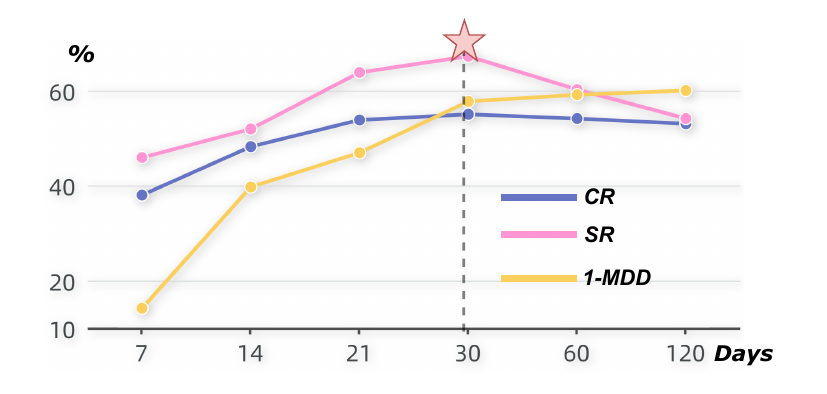}
    \caption{Impact of varying the maximum timescale of the multi-timescale reward on performance metrics.}
    \label{fig:timescale_abla}
\end{figure}

\subsubsection{Multi-Timescale Reward (MTR)}

As shown in Tab.~\ref{tab:ablation_single}, removing MTR causes a severe performance drop across all assets: the CR fall below 20\% for all three stocks (vs. 30.4–62.2\% in the full model), indicating that MTR is  the architectural backbone that enables long-horizon reasoning and strategic consistency in FinPos.

The effectiveness of MTR depends critically on the choice of timescale for computing the cumulative reward. Our design incorporates three horizons (1, 7, and 30 days) to capture immediate fluctuations, medium-term trends, and long-term stability, respectively. We conduct a sensitivity study by varying the length of the long-term horizon (Fig.~\ref{fig:timescale_abla}), finding that: very short windows (7–14 days) yield poor performance, as the agent overreacts to noise without forming stable strategies; performance peaks at 30 days; beyond 30 days, performance declines due to signal dilution, which weakens the reflection mechanism and reduces adaptiveness to regime shifts.

\subsubsection{Quantity and Risk Decision Agent (QRA)}
Removing QRA leads to a significant degradation in risk control. As shown in Tab.~\ref{tab:ablation_single}, while CR drop moderately (e.g., from 62.2\% to 53.6\% on TSLA), the MDD worsens dramatically—increasing from 42.3\% to 62.7\% on TSLA (a 48\% relative increase) and from 27.5\% to 39.3\% on AAPL. This reveals a critical limitation: even with long-horizon awareness provided by MTR, an agent restricted to fixed-unit trades cannot timely adjust its exposure in response to market volatility. Consequently, it lacks the agility needed to hedge positions during sharp, fast-moving drawdowns—precisely the scenarios where position-aware sizing is most valuable.

\subsubsection{Market Signal Processing (MSP)}
Removing MSP leads to moderate performance degradation (e.g., on TSLA: CR drops from 62.2\% to 58.3\%, and MDD worsens from 42.3\% to 45.4\%; see Tab.~\ref{tab:ablation_single}), revealing that raw market data contain substantial noise that distracts downstream agents.
Without MSP’s relevance filtering, an agent may erroneously treat Elon Musk’s tweets about Dogecoin as material signals for TSLA trading, despite their lack of fundamental relevance. Such off-topic social media noise is common in financial news feeds and can trigger spurious trading responses. In contrast, MSP leverages domain-guided heuristics and entity-relevance rules to pre-filter inputs, ensuring that only high-signal information—such as regulatory filings, or company-specific announcements—reaches the LLM analysis pipeline.

\subsection{Risk Analysis}

\begin{figure}[t]
\includegraphics[width=0.49\textwidth]{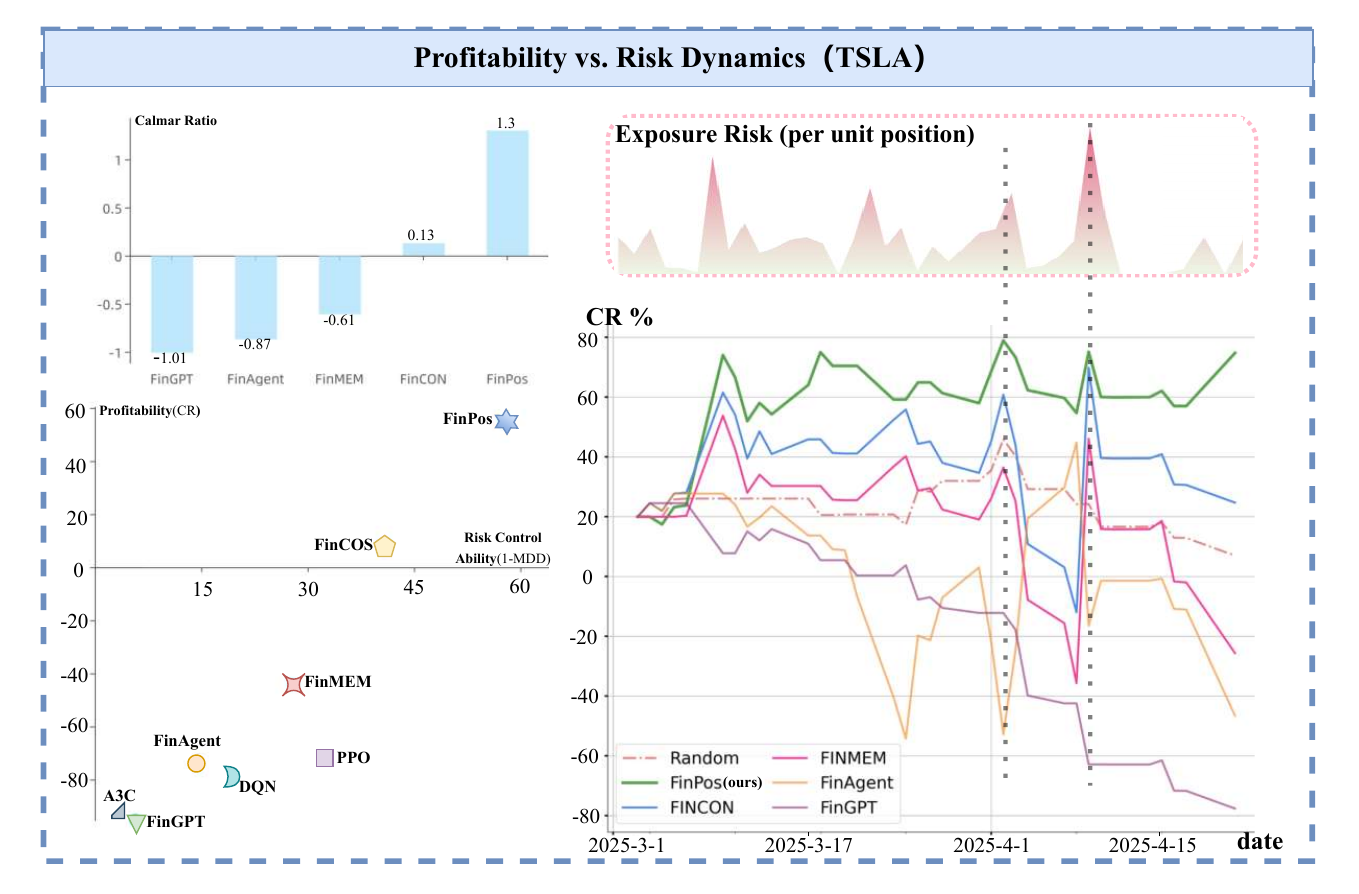}
\caption{Risk-adjusted performance and exposure dynamics on TSLA during Mar–Apr 2025, highlighting FinPos’s advantage under high-volatility events.} 
\label{fig:PA}
\end{figure}

Beyond numerical results, Fig.~\ref{fig:PA} provides focused evidence of FinPos’s risk-adjusted performance during a high-volatility period (Mar–Apr 2025), chosen to highlight its ability to manage extreme market conditions—including those triggered by major political events such as the U.S. election cycle. This window offers a stringent test for position-awareness, as agents must adapt rapidly to shifting risk regimes without overreacting.
The top-left panel reports the calmar ratio (Appendix~\ref{appendix:cr}), which measures return per unit of maximum drawdown. FinPos achieves a calmar of \textbf{1.5}, significantly outperforming all baselines, indicating that its profitability is not driven by excessive risk-taking but by effective control. The bottom-left panel plots cumulative return (CR\%) against risk-control ability (1-MDD). FinPos occupies the upper-right frontier—high return with low drawdown—while baselines cluster in regions of either low return or poor risk management. On the right, the time-series curves and exposure-risk overlay reveal the operational advantage of position-aware management: during high-volatility events (marked by vertical dashed lines), non-PA agents exhibit sharp exposure spikes followed by severe drawdowns, while FinPos proactively reduces exposure and stabilizes returns. This demonstrates that position awareness enables not only reactive mitigation but also anticipatory adjustment—critical for navigating real-world market turbulence.
Additional analyses for AAPL are provided in Appendix~\ref{appendix:aapl}.

\vspace{-0.5em}

%% file: sec/6_Conclusion.tex
\section{Conclusion}

In this paper, we introduce a position-aware trading task that more closely resembles real market conditions compared to the single-step trading task previously employed by researchers. To address this task, we design FinPos, a novel LLM Trading Agent equipped with position-awareness and risk management capabilities. To meet the demands of position management, an LLM agent must possess heightened market sensitivity, enhanced risk control abilities, and a longer-term perspective. Consequently, FinPos employs a specialized market perception module, a dual-decision agent system, and a multi-timescale reward design to achieve these objectives. Our experiments demonstrate that the position management task imposes higher comprehensive ability requirements on LLM Agents. FinPos explores solutions for trading agents operating in this more realistic market environment. 
%However, we also identify two limitations of FinPos that require further attention. Firstly, FinPos is unable to implement coordinated management of multiple asset allocations, thus lacking the capability for higher-level operations such as risk hedging. Secondly, FinPos has been tested only on stock-type investment assets, and its applicability to other asset classes such as futures and bonds remains to be validated.

%% file: sec/7_Limitation.tex
\section*{Limitations}

This work is intended for research purposes only.
Deploying LLM-based trading systems in real-world financial markets without professional oversight
may lead to financial losses, particularly under extreme or unexpected market conditions.

\subsection*{Single-Asset Focus}
Although the proposed framework is general, in this paper we adopt a single-asset trading task as the experimental vehicle to validate its effectiveness.Extending the framework to portfolio-based methods and multi-asset allocation settings is an important direction for future work. 
This will involve addressing cross-asset correlation and rebalancing, their diversified risks, and portfolio-level dynamics, which presents additional complexities that require further model refinement.

\subsection*{Dependency on Prompt Quality:} 
major limitation of FinPOS lies in its sensitivity to prompt design. In our experiments, we focus on well-established companies with abundant and reliable textual information to ensure stable decision-making. As a result, the system relies on asset-specific prompt configurations that are tailored to the characteristics of individual stocks. This design choice limits the direct applicability of FinPOS to newly listed firms or assets with sparse or highly noisy textual signals, where prompt tuning and data curation become necessary. More broadly, this highlights a structural challenge of LLM-based trading agents: their performance is closely tied to the availability and organization of domain-specific textual inputs, rather than solely to model capacity. Addressing this limitation may ultimately require moving beyond prompt-level engineering.
However, model-level adaptation or domain specialization typically demands substantial domain-specific data and computational resources, which we leave to future work.

\subsection*{Reward Design and Multi-Objective Optimization:}
In this work, we adopt a lightweight multi-timescale reward design to regulate position risk and trend alignment, prioritizing stability, interpretability, and reproducibility.
While reinforcement learning (RL) has shown promise for LLM-based agents, its application in financial markets is challenged by non-stationarity, heavy-tailed risks, and regime shifts, which can lead to overfitting and unstable behavior.
Exploring principled RL formulations for learning general financial decision principles remains an important direction for future work.

%% file: sec/8_Appendix.tex
\clearpage

\section{The implementation details of FinPos}

\label{appendix:agent_prompts}
% 简单说几句,然后将prompt都贴上
To provide a clearer understanding of how FinPos operates in practice, we include the complete prompt designs used by each module. These prompts define how the system processes financial information, reasons about market dynamics, and executes trading decisions in a structured manner.

\subsection{Market Signal Processing and Analysis Module}
This module provides the detailed prompts used in the Market Signal Processing and Analysis Module. Each category of textual data (financial reports, quarterly reports, company news, and macroeconomic news) is associated with two prompt templates: one for extracting sentiment and insight, and the other for assessing risk and potential market impact.

% 下面是prompt格式的例子（出自我之前的论文）。注意这种格式不会有tab.1这种标识,是插入到文本之间的。
\lstset{
    % breakindent=0pt,
    framesep = 20pt,
    rulesep = 10pt,
    % breakautoindent=false,
    backgroundcolor = \color[RGB]{245,245,244},
    breaklines = true,
    breakindent = 0pt,
    basicstyle = \ttfamily\small,
    escapeinside = {(*@}{@*)} % 设置 escapeinside 选项
}

\subsubsection{10-K Filings (Annual Reports)}
This section presents the prompts used to process and analyze annual 10-K reports. 
The first prompt extracts and filters key financial information, while the second evaluates its implications for the company's stock price.

\begin{lstlisting}
(*@\color{red}{System Prompt}@*): You are a professional financial analyst. Your task is to evaluate the potential impact of the key points extracted from a company's 10-K report on its stock price, providing decision support for an intelligent trading agent.
You have received the following 10-K key points about {symbol}:
"{filtered_key_points}"
Target company: {symbol}
Please complete the following tasks:
1. Screen and extract the most critical financial and operational highlights, focusing on major changes, performance deviations, strategic shifts, and newly emerging risks or opportunities.
2. Filter out repetitive, boilerplate, or investor-irrelevant information.
3. Rank the retained key points by their importance to investors.
response_format_prompt = Please return the result in the following JSON format, without adding any other explanation:
{
 "key_points": "Selected key points, sorted by importance."
 "reason": "An explanation of why these key points were retained and the rationale behind their importance over other content."
 }
\end{lstlisting} 

\begin{lstlisting}
(*@\color{red}{System Prompt}@*): You are a professional financial analyst specializing in evaluating the short-term and medium-to-long-term impacts of company 10-K reports on stock prices. You are providing decision support for an intelligent trading agent.
You have received a summary of the annual 10-K report for {symbol}:
"{agent_scratch}"
Target company: {symbol}
1. Analyze the potential **short-term (days to one week)** and **medium-to-long-term (weeks to months)** effects on {symbol}'s stock price. Consider whether the developments are likely to surprise the market **positively or negatively** based on typical investor expectations and sentiment.
2. For each impact direction, **differentiate** between contributing factors (e.g., profitability, cash flow, capital allocation, competitive positioning, regulatory risk). Analyze **interactions or trade-offs** between opposing forces.
3. Explain your reasoning in a structured, multi-dimensional way. Go beyond summarization-**synthesize the data**, explore **counterfactual scenarios**, and account for **macro and industry context**. If relevant, mention **investor psychology** or narrative shifts.
4. DO NOT simply restate the report. Your goal is to **interpret, evaluate, and draw meaningful implications** for trading behavior and valuation outlook.
Please maintain professionalism, clarity, and logical coherence. Highlight key opportunities and risks with balanced and nuanced judgment.
Please output only the response in JSON format without any additional commentary:
{response_format_prompt}= Please return the result in the following JSON format, without adding any other explanation:
{
  "insight": "This 10-K report is positive/negative/neutral for {symbol} in the short term, and positive/negative/neutral in the medium to long term.",
  "reason": "Explain the core reasoning behind the judgment, reflecting logical analysis of the key points."
}
\end{lstlisting}

\subsubsection{10-Q Filings (Quarterly Reports)}
Similarly, the following prompts are designed for 10-Q reports, focusing on quarterly performance updates and market reactions.

\begin{lstlisting}
(*@\color{red}{System Prompt}@*): You are a professional financial analyst. Your task is to evaluate the potential impact of the key points extracted from a company's 10-Q report on its stock price, providing decision support for an intelligent trading agent.
You have received the following 10-Q key points about {symbol}:
"{filtered_key_points}"
Target company: {symbol}
Please complete the following tasks:
1. Screen and extract the most critical financial and operational highlights, focusing on major changes, performance deviations, strategic shifts, and newly emerging risks or opportunities.
2. Filter out repetitive, boilerplate, or investor-irrelevant information.
3. Rank the retained key points by their importance to investors.
response_format_prompt = Please return the result in the following JSON format, without adding any other explanation:
{
 "key_points": "Selected key points, sorted by importance."
 "reason": "An explanation of why these key points were retained and the rationale behind their importance over other content."
 }
\end{lstlisting} 

\begin{lstlisting}
(*@\color{red}{System Prompt}@*): You are a professional financial analyst specializing in evaluating the short-term and medium-to-long-term impacts of company 10-Q reports on stock prices. You are providing decision support for an intelligent trading agent.
You have received a summary of the quarterly 10-Q report for {symbol}:
"{agent_scratch}"
Target company: {symbol}
1. Analyze the potential **short-term (days to one week)** and **medium-to-long-term (weeks to months)** effects on {symbol}'s stock price. Consider whether the developments are likely to surprise the market **positively or negatively** based on typical investor expectations and sentiment.
2. For each impact direction, **differentiate** between contributing factors (e.g., profitability, cash flow, capital allocation, competitive positioning, regulatory risk). Analyze **interactions or trade-offs** between opposing forces.
3. Explain your reasoning in a structured, multi-dimensional way. Go beyond summarization-**synthesize the data**, explore **counterfactual scenarios**, and account for **macro and industry context**. If relevant, mention **investor psychology** or narrative shifts.
4. DO NOT simply restate the report. Your goal is to **interpret, evaluate, and draw meaningful implications** for trading behavior and valuation outlook.
Please maintain professionalism, clarity, and logical coherence. Highlight key opportunities and risks with balanced and nuanced judgment.
Please output only the response in JSON format without any additional commentary:
{response_format_prompt}
response_format_prompt = """Please respond in the following JSON format **without adding any additional explanations**:
{
  "key_points": "Concise summary of the most critical content, including key highlights and risks with brief explanation.",
  "insight": "This report has a positive/negative/neutral impact on {symbol} in the short term, and a positive/negative/neutral impact in the medium to long term.",
  "reason": "Comprehensive explanation of the reasoning behind the judgment, showing multi-dimensional logical analysis and complex factor consideration rather than simple summary."
}
"""
\end{lstlisting}

\subsubsection{Macroeconomic News}
This section presents the prompts used to process macroeconomic news and policy releases. Two corresponding prompt templates are adopted: the first filters and ranks macroeconomic news items by their relevance and significance to thethe target company, while the second analyzes how these events may influence investor sentiment, capital flows, and asset price movements over short and medium-to-long horizons.

\begin{lstlisting}
(*@\color{red}{System Prompt}@*): You are an experienced financial research assistant. Your task is to determine whether a given news article is related to a specific company.
Target company: {symbol}
Please analyze the news below and classify the relationship between the news and the company as either "direct", "indirect", or "none", according to the criteria provided:

Classification criteria:
1. If the news **explicitly mentions** the company name (e.g., Tesla), its executives (e.g., Elon Musk), its products, financial reports, mergers, partnerships, or investments -- classify as **direct**
2. If the news does **not explicitly mention** the company, but includes topics that **have a substantial impact on the company's business, valuation, or market performance**  -- classify as **indirect**, such as:
   - Industry level: industry trends, changes in market demand, technological advancements, industry regulatory policies, upstream/downstream supply chain, competitor dynamics, price fluctuations of key materials (e.g., lithium, batteries)
   - Macroeconomic factors: macroeconomy, Federal Reserve policies, interest rates, inflation, employment, consumer spending, GDP growth, manufacturing indices, PMI, retail sales, and other macroeconomic indicators
   - Financial market sentiment: significant fluctuations or sustained trends in the S&P 500, Nasdaq, Dow Jones indices; market overheating, overbought conditions, or panic selling that may affect overall risk appetite; valuation adjustments in tech/growth stock sectors; market rotation; financing environment; IPO activities; large ETF inflows or outflows
   - Policies and regulations: national policies, taxation, regulation, energy, climate, green transition, green energy subsidies, emission standards, electric vehicle regulations; US-China trade war, export restrictions, chip bans, customs policies, etc.
   - US-China relations, export controls, trade wars, tariff adjustments, technology bans
   - Geopolitical conflicts (e.g., Russia-Ukraine war, Middle East tensions), international sanctions, energy price surges causing global market volatility or disruptions in energy/logistics/supply chains
   - Key figures (e.g., Trump, Biden, Federal Reserve Chair Powell) making political, economic, or policy statements, policy preferences, election outlooks, trade comments, or antitrust remarks
3. If the news is **not substantively related** to the company and is **unlikely to impact** its operations or stock price -- classify as **none**, such as:
   - Natural disasters, entertainment gossip, or local events unrelated to the company's business, industry, or market
   - Regional incidents with no significant impact on the company's country's economy or policies

News article:
{agent_scratch}
Please output only the response in JSON format without any additional commentary:
{response_format_prompt}

response_format_prompt = """Please output a JSON in the following format:
{
  "relation_type": "direct" | "indirect" | "none",
  "reason": "Briefly explain the reasoning behind your judgment"
}
"""
\end{lstlisting}

\begin{lstlisting}
(*@\color{red}{System Prompt}@*): You are a professional financial analyst specializing in evaluating the medium- to long-term impact of financial news on company stock prices. You are assisting an intelligent trading agent with decision-making support.
You have received the following financial news:
"{agent_scratch}"
The target company is: {symbol}

Please complete the following tasks:
1. Do **not** repeat or summarize the original news content;
2. Determine whether this news has a **material impact** on {symbol}'s stock price, not limited to direct relevance - please also consider macroeconomic policy, supply chain dynamics, market sentiment, geopolitical risks, or other indirect or lagging factors;
3. If there is an impact, provide **one clear and concise investment insight**, explaining how the news might affect {symbol}'s stock price in the coming **weeks to months** (e.g., bullish or bearish);
4. If there is **no clear relevance or impact**, clearly state that the news has **no significant effect** on {symbol};
5. Evaluate the relevance level of the news to {symbol}, using the following scale:
    "high": The news has a direct and significant impact on the company's fundamentals, financials, regulatory environment, or industry position;
    "medium": The news could have an indirect or delayed impact, such as through macroeconomic trends, industry supply/demand shifts, investor sentiment, or cost structure changes;
    "low": The news is largely unrelated or only remotely connected to the company.
Please output only the response in JSON format without any additional commentary:
{response_format_prompt}

response_format_prompt = """Please respond using the following JSON format and do not include any additional text:
{
  "insight": "Summary of how this news may impact {symbol}",
  "relevance": "high" | "medium" | "low"
}
"""
\end{lstlisting}

\subsubsection{company news}
Since company-specific news can be directly collected by ticker symbol, the filtering process focuses primarily on assessing **relevance, materiality, and potential market impact**, rather than broad topic association. 
Two prompt templates are used in this module - one for filtering and ranking important news items, and the other for analyzing their short-term and medium-to-long-term effects on stock performance.

\begin{lstlisting}
(*@\color{red}{System Prompt}@*): You are a professional financial analyst. Your task is to filter and prioritize firm-level news items based on their potential importance to investors and their relevance to the company's stock price.
You have received several pieces of company-related news for {symbol}:
"{news_batch}"

Please complete the following steps:
1. Identify which items are **material** and likely to influence investor perception or price movement;
2. Filter out minor, repetitive, or purely descriptive updates with limited market relevance;
3. Rank the retained items by their expected significance to the stock price, considering tone, topic, and potential investor reaction.

Please return the result strictly in JSON format:
{
  "key_points": "Selected and ranked company news items that are most likely to affect {symbol}'s stock price.",
  "reason": "Explain briefly why these items are more significant than others."
}
\end{lstlisting}

\begin{lstlisting}
(*@\color{red}{System Prompt}@*): You are a professional financial analyst specializing in assessing the **price sensitivity** of company-related news. You are assisting a high-performance trading agent that only acts based on material, relevant information.

Here is a piece of news you received:
"{agent_scratch}"
Target company: {symbol}

Please follow these instructions:
1. Do NOT summarize the news content;
2. Focus ONLY on the potential impact of this news on {symbol}'s stock price;
3. If this news is irrelevant or has no clear directional impact on {symbol}, clearly mark it as **"neutral"** with an appropriate reason;
4. Evaluate the likely impact in both:
   - **Short term** (1-5 trading days)
   - **Medium to long term** (a few weeks to months);
5. Be strict: only assign "positive" or "negative" if the news provides clear evidence of directional influence on {symbol}'s fundamentals or investor sentiment.

Please output only the response in JSON format without any additional commentary:
response_format_prompt = """Please return the result in the following **JSON format**, without adding any extra explanation:
{
  "insight": "This news has a [positive/negative/neutral] impact on {symbol} in the short term, and a [positive/negative/neutral] impact in the medium to long term.",
  "reason": "Explain the key reasoning behind your assessment. Do not summarize the news content."
}
"""
\end{lstlisting}

\subsection{Dual Trading Decision Module}

This section provides the detailed prompt structures used in the Dual Trading Decision Module. While the architectural overview (see Fig.~\ref{fig:archi}) already explains the interaction flow, here we focus on the internal prompt logic and reasoning objectives of each decision agent.

\subsubsection{Direction Decision Agent}
The following prompt guides the agent to leverage the key insights extracted by the preceding analytical modules and determine the optimal trading direction (buy, sell, or hold), along with the overall strategic orientation for the current trade.
\begin{lstlisting}
(*@\color{red}{System Prompt}@*): 
# memory IDs 
short_memory_id_desc = "ID of short-term information."
mid_memory_id_desc = "ID of mid-term information."
long_memory_id_desc = "ID of long-term information."
reflection_memory_id_desc = "ID of reflective-period information."
train_memory_id_extract_prompt = "Select and store the most investment-relevant information from major sources (e.g., ARK, Two Sigma, Bridgewater Associates) into the {memory_layer} memory."
test_memory_id_extract_prompt = "Retrieve the most relevant information from the {memory_layer} memory for the current investment decision."

# trading summary 
train_trade_reason_summary = "Based on a professional trader's advice, explain why the trader would make such a decision given the provided information."
test_trade_reason_summary = "Based on the text information and summarized price trends, explain the reason for your investment decision."
test_invest_action_choice = "Based on the information, choose one of the following actions: buy, sell, or hold."

# investment info 
train_investment_info_prefix = (
    "The current date is {cur_date}. The observed market facts are as follows: "
    "For {symbol}, the price difference between the next and current trading day is {cur_record_t1}; "
    "the 7-day difference is {cur_record_t7}; "
    "the 30-day difference is {cur_record_t30}. "
    "Your decision return is {reward}.\n\n"
)
test_investment_info_prefix = "The stock under analysis is {symbol}, and the current date is {cur_date}."

# sentiment & momentum explanation 
test_sentiment_explanation = """For example, positive news about a company may boost investor confidence and trigger buying activities, pushing the stock price upward;
whereas negative news tends to dampen sentiment, leading to selling pressure and price declines.
Additionally, news related to competitors or the broader industry can indirectly affect the target stock's performance.
Sentiment scores (positive, neutral, negative) represent the distribution across these categories (summing to 1) and, together with "importance" and "timeliness" indicators, help assess the market impact and validity of the information.
"""
test_momentum_explanation = """The following summarizes recent price movements, i.e., momentum.
Momentum reflects the idea that stocks performing strongly in the short term often continue rising,
while weak performers are more likely to keep declining.
"""

# training phase prompt 
train_prompt = """Please complete the following two tasks based on the investment information below:
Important: Do NOT use any future price differences (T+1, T+7, T+30) in your reasoning. These are unavailable in real-time trading. Any output referencing them will be considered invalid.
1. Directional Decision:
Choose one of the following actions: "buy", "sell", or "hold" (only if uncertain).
You must consider:
- Information from short-, mid-, long-term, and reflective memories;
- Historical price momentum;
- Sentiment tendencies, importance, and timeliness in news or reports.
Briefly describe your decision logic, the overall trading strategy (e.g., long-term accumulation or short-term profit), and indicate the supporting memory indices.
2. Reflection:
The system will automatically evaluate whether your directional judgment matches the market trend.
- If incorrect, explain the misinterpreted or overemphasized information.
- If correct, summarize the key factors behind the correct judgment.
${investment_info}

Your output must strictly follow the JSON format below, with no extra text:
{
    "investment_decision": "buy" | "sell" | "hold",
    "summary_reason": "Brief explanation of your decision logic",
    "short_memory_index": [integer list],
    "middle_memory_index": [integer list],
    "long_memory_index": [integer list],
    "reflection_memory_index": [integer list],
    "reflection_analysis": "Reflection analysis text"
}
"""

# testing phase prompt 
test_prompt = """Determine the optimal investment direction based on the following information and briefly justify your reasoning.
You must consider:
- Information from all memory layers (short-, mid-, long-term, reflective);
- Historical price momentum;
- The importance, sentiment, and timeliness of key information.
Provide one of three decisions: "buy", "sell", or "hold", and indicate the memory IDs supporting your judgment.

${investment_info}
${gr.complete_json_suffix_v2} }
"""
\end{lstlisting}

\subsubsection{Quantity and Risk Decision Agent}
The following prompt instructs the agent to determine the specific order quantity for the current trade based on the analytical results and strategic guidance from the Direction Decision Agent, while adjusting for current holdings and potential risk exposure.
\begin{lstlisting}
(*@\color{red}{System Prompt}@*): 
# memory IDs 
short_memory_id_desc = "ID of short-term information."
mid_memory_id_desc = "ID of mid-term information."
long_memory_id_desc = "ID of long-term information."
reflection_memory_id_desc = "ID of reflective-period information."
train_memory_id_extract_prompt = "Select and store the most investment-relevant information from major sources (e.g., ARK, Two Sigma, Bridgewater Associates) into the {memory_layer} memory."
test_memory_id_extract_prompt = "Retrieve the most relevant information from the {memory_layer} memory for the current investment decision."

# trading summary 
train_trade_reason_summary = "Based on a professional trader's advice, explain why the trader would make such a decision given the provided information."
test_trade_reason_summary = "Based on the text information and summarized price trends, explain the reason for your investment decision."
test_invest_action_choice = "Based on the information, choose one of the following actions: buy, sell, or hold."

# investment info 
train_investment_info_prefix = (
    "The current date is {cur_date}. The observed market facts are as follows: "
    "For {symbol}, the price difference between the next and current trading day is {cur_record_t1}; "
    "the 7-day difference is {cur_record_t7}; "
    "the 30-day difference is {cur_record_t30}. "
    "Your decision return is {reward}.\n\n"
)
train_reward_explanation =  """Reward reflects the quality of your past decision:
        - **Positive**: Good decision; higher means better alignment with market.
        - **Negative **: - **Negative**: A weaker decision. The more negative the value, the worse the outcome - may caused by misreading available data.
        Use reward **only for reflection**, not for future predictions.
"""
test_investment_info_prefix = "The stock under analysis is {symbol}, and the current date is {cur_date}."

# sentiment & momentum explanation 
test_sentiment_explanation = """For example, positive news about a company typically boosts market confidence, stimulates buying, and drives up the stock price;
Negative news, on the other hand, weakens confidence, triggers selling pressure, and causes the stock price to fall.
Industry or competitor dynamics may also indirectly affect the target company's performance.
The sentiment score (positive, neutral, negative) reflects the proportion of the text in each of the three sentiment categories (summing to 1).
It can be combined with the "Importance" and "Timeliness" metrics to assess the market impact and validity of the information.
In addition, you need to combine the output of the previous Direction Decision Agent (i.e., the overall strategic description of this transaction) as the strategic basis for quantitative decisions.
"""
test_momentum_explanation = """The following summarizes recent price movements, i.e., momentum.
Momentum reflects the idea that stocks performing strongly in the short term often continue rising,
while weak performers are more likely to keep declining.
"""

# training phase prompt 
train_prompt = """Please complete the following two tasks based on the investment information below:
Important: Do NOT use any future price differences (T+1, T+7, T+30) in your reasoning. These are unavailable in real-time trading. Any output referencing them will be considered invalid.
1. Investment Amount and Risk Decision:
You already know the directional decision (buy/sell/hold) made by the Direction Decision Agent in the previous stage.
Based on this, determine the **specific order quantity**(integer) and ensure that the transaction volume does not exceed the maximum limit {maxcvar} recommended by the risk control module.
You must consider the following factors:
- Information in each memory layer (short-term, medium-term, long-term, and reflection period);
- Historical momentum and price volatility;
- The sentiment, importance, and timeliness of news or financial reports;
- Current account holdings and overall risk exposure;
- Trading strategy determined in the previous phase.
Please briefly explain your quantity decision logic and indicate the memory indexes supporting this decision.
2. Decision Reflection and Analysis:
The system will calculate a reward based on the order quantity and corresponding return.
- If the reward is negative, please explain any market signals or risk factors you may have misjudged;
- If the reward is positive, please summarize the core rationale that led to your correct decision.
${investment_info}

Your output should strictly adhere to the following JSON format and not include any other content:
{
"order_size": integer (range 1 to {maxcvar}),
"summary_reason": "Please enter your quantity and risk decision logic here",
"short_memory_index": [list of integers],
"middle_memory_index": [list of integers],
"long_memory_index": [list of integers],
"reflection_memory_index": [list of integers],
"reflection_analysis": "Please fill in your reflection description here."
}
"""

# Testing phase prompt
test_prompt = """Based on the following information, please determine the **order quantity** for the current trade.
You know the directional decision (buy/sell/hold). Please specify the specific order quantity based on the risk exposure and CVaR constraint (maximum order quantity {maxcvar}).
You must consider:
- Memory information at each level (short-term, medium-term, long-term, reflection period);
- Momentum trend, sentiment, information importance, and timeliness;
- Current account holdings and overall risk;
- Trading strategy for the previous directional decision.
Please output a specific order quantity (integer, not exceeding {maxcvar}) and indicate the information index that supports your judgment.

${investment_info}
${gr.complete_json_suffix_v2} }
"""
\end{lstlisting}

\section{Formulas of Classic Financial Metrics}

To evaluate the risk-return characteristics of trading strategies, we summarize the formal definitions of commonly used financial evaluation metrics, including risk-adjusted return and downside risk measures, which are used throughout our experiments.

\subsection{Definitions of Evaluation Metrics}
\label{app:metrics}

\begin{equation}
\text{Sharpe Ratio} = \frac{R_p - R_f}{\sigma_p}, \quad
\end{equation}
where $R_p$ is the average return of the portfolio, $R_f$ is the risk-free rate, and $\sigma_p$ is the standard deviation of portfolio returns. A higher sharpe ratio indicates more efficient risk-adjusted performance.

\begin{equation}
\text{MDD} = \max_{t=1}^N\left(\frac{ P^{t} - P_{trough}^{t}}{ P^{t}}\right)
\end{equation}
where $t$ denotes the index of the trading day, $P^{t}$ is the account value (the market value of the current stock position) at day $t$, and $P_{trough}^{t}$ is the lowest future account value observed after day $t$. A smaller MDD reflects stronger downside protection and greater robustness of the strategy.

\subsection{Conditional Value at Risk (CVaR)}
\label{appendix:cvar}

Let the profit and loss over a trading horizon be denoted by \( PnL \).  
The Value at Risk (VaR) at a confidence level \( \alpha \) represents the maximum potential loss not exceeded with probability \( \alpha \), formally defined as:
\[
\text{VaR}_{\alpha}(PnL) = \inf \{ l \in \mathbb{R} : P(PnL \le l) \ge \alpha \}.
\]
The Conditional Value at Risk (CVaR) measures the expected loss that occurs beyond the VaR threshold, providing a more comprehensive view of downside risk:
\[
\text{CVaR}_{\alpha}(PnL) = \mathbb{E}[PnL \mid PnL \le \text{VaR}_{\alpha}(PnL)].
\]
Importantly, during trading, CVaR is computed online using a 20-trading-day rolling window of past realized returns and updated daily, ensuring that no forward-looking or test-period information is used in position sizing.  A smaller CVaR indicates stronger downside protection and more effective risk control.

\subsection{Calmar Ratio}
\label{appendix:cr}

The Calmar Ratio evaluates the trade-off between return and maximum drawdown. It is defined as:
\[
\text{Calmar Ratio} = \frac{R_{\text{annual}}}{|\text{MDD}|}
\]
where $R_{\text{annual}}$ denotes the annualized return, and $\text{MDD}$ represents the maximum drawdown during the same period.  
A higher Calmar Ratio indicates a better risk-adjusted performance.

\section{Experimental Setup}
\label{appendix:es}

\subsection{Position Awareness and Decision Structures}

As summarized in Tab.~\ref{tab:llm-based}, most existing LLM-based trading agents formulate decision-making as a discrete action selection problem, typically restricted to \emph{buy}, \emph{sell}, or \emph{hold}, with fixed or implicit trade sizes. FinGPT adopts a single-agent architecture with predetermined position sizes, while FinMem leverages memory mechanisms to guide reasoning without explicitly modeling position magnitude or exposure. Although FinCon and FinAgent employ multi-stage or debate-based reasoning pipelines, their final outputs remain direction-only decisions without explicit position sizing.
However, trading volume is a fundamental component of position awareness. Without the ability to adjust exposure magnitude, agents operating under direction-only actions lack fine-grained control over risk. For instance, when holding a large long position, an agent may detect increased downside risk but can only respond through unit-based actions, resulting in delayed or insufficient risk mitigation.
In contrast, FinPos explicitly models position evolution through a two-stage decision structure that decouples directional reasoning from quantitative position sizing. By incorporating CVaR-based position control and reflection guided by multi-timescale rewards, FinPos enables risk-aware adjustments to both trading direction and exposure magnitude, leading to more coherent and realistic position management.

\subsection{LLM-based Baselines Trading Agents}

We compare FinPos against a representative set of state-of-the-art LLM-based trading agents that differ in architectural design, information processing pipelines, and decision structures, as summarized in Tab~\ref{tab:llm-based}.
We strictly follow the inference settings reported in their original papers.

\begin{table*}[ht]
\centering
\caption{Overview of LLM-based Trading Agents and Their Architectures}
\label{tab:llm-based}
\resizebox{\textwidth}{!}{ % 让表格宽度适应页面
\begin{tabular}{|l|l|p{3.5cm}|p{3cm}|p{4cm}|p{5cm}|}
\hline
\textbf{Method} & \textbf{Backbone} & \textbf{Agents / Modules} & \textbf{Position-Aware} & \textbf{Information Processed} & \textbf{Decision Structure} \\
\hline
FinPOS     & GPT-4o  & 10 agents: 4 filters + 4 analyzers + Direction Agent + Quantity/Risk Agent & Yes (continuous position state + CVaR sizing) & Price, news, macro events, 10-K/10-Q reports & Two-stage: (1) direction reasoning; (2) CVaR-based position sizing; reflection with multi-timescale reward \\
\hline
FinCon     & GPT-4o  & 7 agents: Data, News, Report, Analyst Report, Earnings Call, Stock Selection, Manager & No (no explicit position sizing; only discrete actions) & Structured \& unstructured financial data & Multistage pipeline with manager aggregation; buy/sell/hold \\
\hline
FinAgent   & GPT-4o  & 10+ agents: Fundamental Analyst, Sentiment Analyst, News Analyst, Technical Analyst, Bullish/Bearish Researchers, Trader, Risk (3 agents), Fund Manager & No (no explicit position sizing; only discrete actions) & Sentiment, technical indicators, news, fundamentals (e.g., revenue, profit, debt) & Debate + risk-check pipeline; buy/sell/hold \\
\hline
FinGPT     & llama (fine-tuned)  & Single agent & No & Sentiment signals + price & Generates buy/sell signals with fixed trade size \\
\hline
FinMem     & GPT-4o  & Single agent & No (memory guides reasoning only) & Data, News, 10-K/10-Q & Single-step; buy/sell/hold \\
\hline
\end{tabular}
}
\end{table*}

\subsection{Deep Reinforcement Learning Baselines}
In addition to LLM-based agents, we benchmark FinPos against classical deep reinforcement learning (DRL) agents, including A2C, PPO, and DQN. All DRL baselines are implemented using the FinRL framework~\cite{Finrl}. These agents operate solely on numerical features derived from market prices and technical indicators, without access to textual or semantic information. The key hyperparameters for each RL agent are listed in Table~\ref{tab:rl_hyperparams}.

\begin{table}[ht]
\centering
\caption{Key Hyperparameters for RL Agents}
\label{tab:rl_hyperparams}
\begin{tabular}{|l|l|l|}
\hline
\textbf{Agent} & \textbf{Key Hyperparameters} & \textbf{Value} \\
\hline
A2C   & n\_steps             & 5     \\
      & ent\_coef            & 0.01  \\
      & learning\_rate       & 0.0007 \\
\hline
PPO   & n\_steps             & 2048  \\
      & ent\_coef            & 0.01  \\
      & learning\_rate       & 0.00025 \\
      & batch\_size          & 64    \\
\hline
DQN   & batch\_size          & 128   \\
      & buffer\_size         & 50000 \\
      & learning\_rate       & 0.001 \\
\hline
\end{tabular}
\end{table}

\begin{figure}[t]
    \centering
    \includegraphics[width=0.9\linewidth]{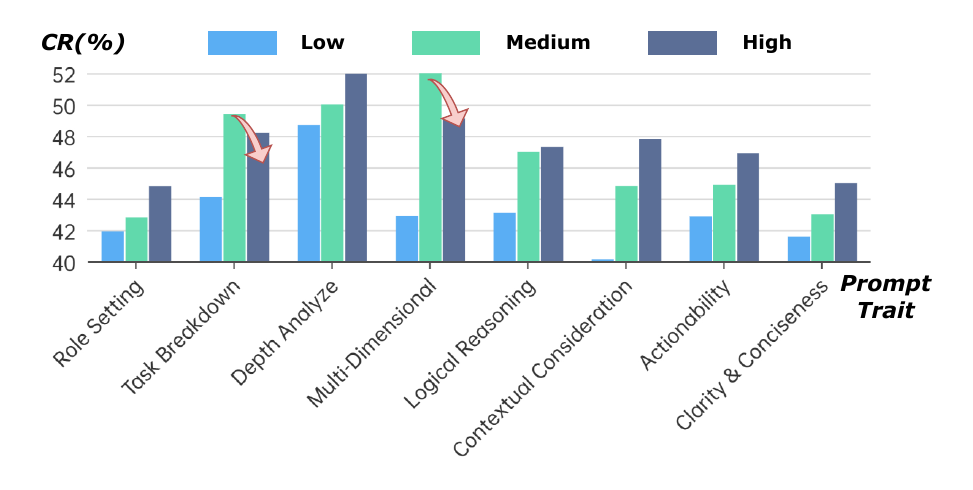}
    \caption{Prompt ablation across eight characteristics and three emphasis levels.}
    \label{fig:prompt_abla}
\end{figure}

\section{More Experiments}

\begin{figure*}[t]
\includegraphics[width=\textwidth]{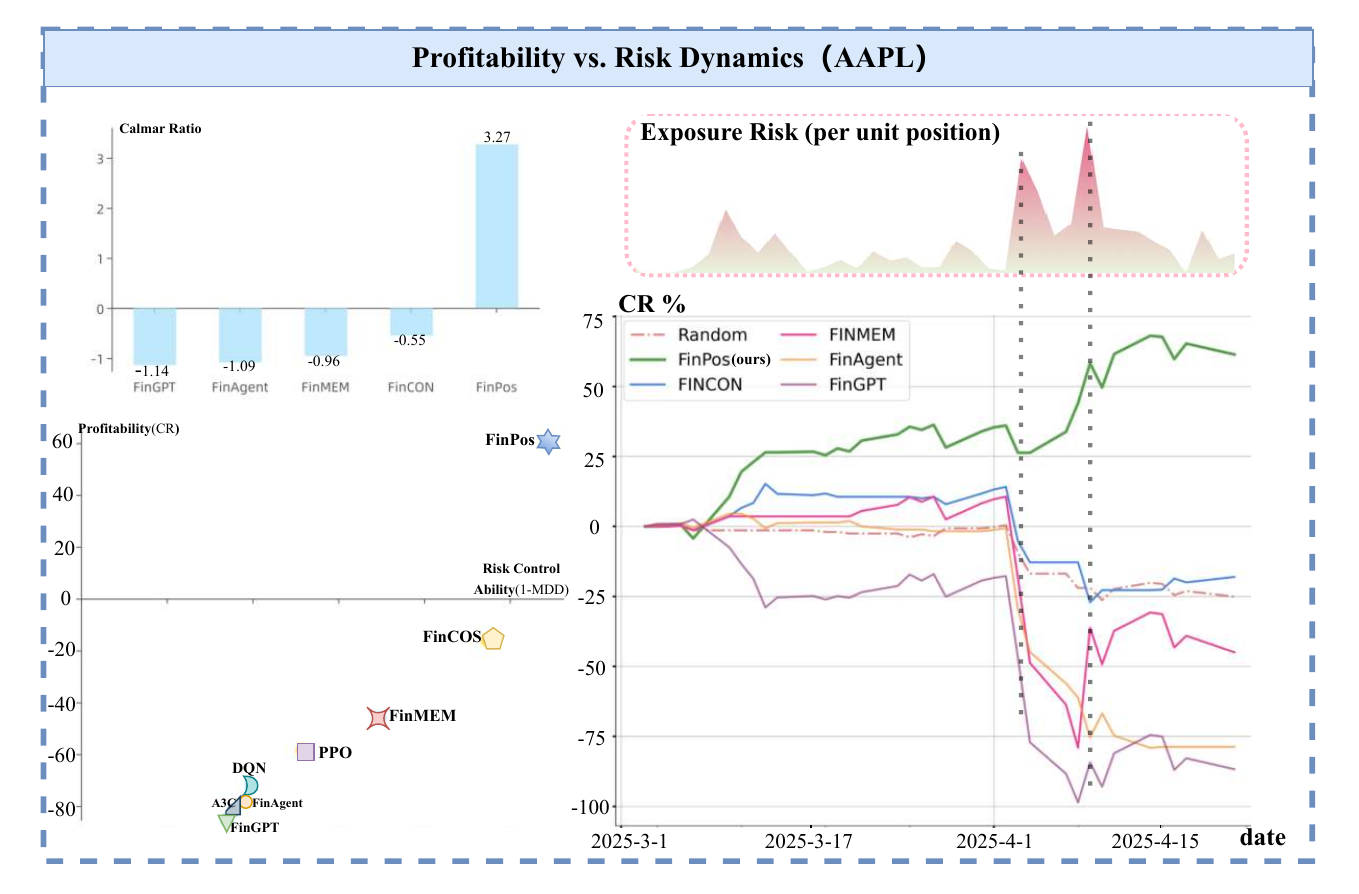}
\caption{Profitability versus risk dynamics. 
Top-left: Calmar ratio, capturing return relative to maximum drawdown (higher is better). 
Bottom-left: joint view of profitability (CR\%) and risk-control (1 - MDD), indicating return-risk balance. 
Right: time-series comparison of cumulative return (bottom) and exposure risk (top) across major events; PA-aware strategies mitigate exposure spikes and sustain more stable growth.} 
\label{fig:PA2}
\end{figure*}

\subsection{Sensitivity to LLM Sampling Hyperparameters}

FinPOS does not train or fine-tune large language models; therefore, its performance is independent of random initialization seeds typically used in neural network training. However, to address reviewers' concerns regarding stochasticity introduced during LLM inference, we conduct a sensitivity analysis by varying key decoding hyperparameters, including temperature and top\_p.

Specifically, we evaluate FinPOS under multiple sampling configurations on TSLA while keeping all other components unchanged. As shown in Tab.~\ref{tab:sampling-sensitivity}, the overall performance remains stable across different settings, with only minor variations in cumulative return (CR), Sharpe ratio (SR), and maximum drawdown (MDD). These results indicate that FinPOS is not overly sensitive to reasonable changes in LLM sampling strategies, and its trading behavior is robust under inference-time stochasticity.

\begin{table}[t]
\centering
\small
\caption{Sensitivity Analysis of FinPOS under Different LLM Sampling Hyperparameters (TSLA)}
\label{tab:sampling-sensitivity}
\begin{tabular}{|p{3cm}|c|c|c|}
\hline
\textbf{Sampling Setting} & \textbf{CR (\%)} & \textbf{SR} & \textbf{MDD (\%)} \\
\hline
Default ($T$=0.7, $p$=0.9) & 62.15 & 0.68 & 42.34 \\
$T$=0.7, $p$=0.85         & 61.74 & 0.67 & 42.50 \\
$T$=0.7, $p$=0.80         & 62.01 & 0.66 & 42.40 \\
$T$=0.8, $p$=0.9          & 60.13 & 0.66 & 45.05 \\
$T$=0.5, $p$=0.9          & 61.07 & 0.64 & 43.10 \\
\hline
\end{tabular}
\end{table}

\subsection{More Stock Trading Result Graphs}
\label{appendix:aapl}
% 模仿FinCON A.6中的形式放8个股票的收益波动图即可,不需要像图5一样放太多复杂的东西。
Similarly, for AAPL, the supplementary plots (Fig.~\ref{fig:PA2}) provide complementary evidence from profitability and risk control. FinPos again occupies the upper-right region in the CR\% vs. risk-control space, indicating that its returns are achieved without compromising drawdown management. The time-series and exposure-risk overlays highlight that, during high-volatility periods, non-position-aware methods experience sharp spikes in exposure and subsequent losses, whereas FinPos effectively anticipates and mitigates these risks. Overall, these results reinforce that position awareness consistently improves risk-adjusted performance across different stocks.

\subsection{More Details of Ablation Studies}
\label{appendix:Ablation}

\subsubsection{Financial Insight Prompting (FIP)}
\label{appendix:fip}
Financial Insight Prompting (FIP): A targeted prompting strategy designed to mitigate LLMs’ weaknesses in financial reasoning. It gradually instills financial thinking by emphasizing causal chains, market trend, and probabilistic inference.

Introducing FIP leads to a clear overall performance gain, consistently enhancing cumulative return across all assets (e.g., TSLA: 52.56\% → 62.15\%; AAPL: 59.38\% → 67.31\%). 
To investigate the impact of prompt design on the depth of agent finance insight, we divide the content of prompts into eight key dimensions: Role Setting, Task Breakdown, Depth Analysis, Multi-Dimensional Analysis, Logical Reasoning, Contextual Consideration, Actionability, and Clarity \& Conciseness(the detailed explanations of these eight dimensions are provided in Tab.~\ref{tab:prompt_traits}). For each dimension, we design prompts with different levels of refinement—Low, Medium, and High—and conduct graded experiments. The results are shown in Fig.~\ref{fig:prompt_abla}.

The results indicate that Task Breakdown, Depth Analysis, and Multi-dimensional Analysis are the most influential dimensions for overall agent performance. Notably, Depth Analysis plays a particularly crucial role in financial reasoning: since the quality of financial analysis is significantly influenced by deep insights, particularly those concerning causal chains and risk-reward tradeoffs.
In contrast, Multi-Dimensional Analysis presents a more nuanced pattern. Financial analysis naturally spans across multiple perspectives, such as macro vs. micro and short-term vs. long-term. While moderate prompting enhances the model’s ability to cover multiple factors, overly elaborate prompts may overwhelm the LLM, causing it to “overthink,” lose focus, and eventually degrade performance—an effect we refer to as the information burden.
Among the eight dimensions, Contextual Consideration shows the most pronounced performance shift. Without explicit prompting, the model tends to overlook this dimension—for instance, it may treat tariff adjustments as entirely unrelated to stock prices. Once guided, however, the agent develops stronger contextual reasoning: it can link major events (e.g., political shifts, regulatory changes) to firm-level responses (e.g., strategic adjustments to tariffs), and further to market outcomes. This ability allows the model to generate insights that align more closely with real-world market dynamics.

Overall, the optimal strategy in FIP design is to selectively emphasize key dimensions—particularly Depth Analysis, Contextual Consideration, and Logical Reasoning—while keeping the prompt concise and focused. Such design not only improves the quality of reasoning but also ensures that the conclusions provide actionable guidance in real financial markets.

\begin{table}[ht]
\centering
\caption{Evaluation Criteria for Prompt Traits}
\label{tab:prompt_traits}
\begin{tabular}{p{0.28\linewidth}|p{0.68\linewidth}}  % 调整宽度使表格跨栏且可读
\hline
\textbf{Trait} & \textbf{Evaluation Criteria (How to measure)} \\ \hline
Role Setting & Does it clearly assign a professional role to the AI (e.g., "Financial Analyst")? \\ \hline
Task Breakdown & Does it break down complex analytical tasks into clear sub-tasks? \\ \hline
Depth Requirement & Does it explicitly ask the model to infer and analyze, not just summarize or restate? \\ \hline
Multi-dimensional Analysis & Does it guide the model to consider multiple factors (e.g., short-term/long-term, macro/micro)? \\ \hline
Logical Reasoning & Does it encourage the model to build causal chains, weigh pros and cons, and infer potential impacts? \\ \hline
Contextual Consideration & Does it guide the model to consider non-financial factors (e.g., investor psychology, market narrative)? \\ \hline
Actionability & Does it require the model to provide conclusions that have practical guidance for decision-making? \\ \hline
Clarity \& Conciseness & Is the prompt itself easy to understand, unambiguous, and not redundant? \\ \hline
% \label{tab:fip}
\end{tabular}
\end{table}

\subsubsection{Signal Ablation across Analyst Agents}

To better isolate the contributions of individual information sources, we conduct an agent-wise signal ablation study by selectively disabling specific analyst agents while keeping all other components unchanged.
This design allows us to quantify the marginal effect of each signal type under identical market conditions.

As this experiment aims to analyze the relative importance of different information sources rather than cross-asset generalization,
we report representative results on TSLA, which offers the richest and most comprehensive set of textual signals.
We observe consistent qualitative trends across other assets.

As shown in Table~\ref{tab:signal-ablation}, removing any individual signal source leads to a noticeable degradation in performance, though the magnitude varies by signal type.
Disabling quarterly filings (10-Q) results in a relatively moderate decline.
In contrast, removing annual filings (10-K) substantially increases maximum drawdown, indicating their importance in anchoring long-term risk perception.
Overall, company-level news has the most pronounced impact on both return and risk metrics, followed by macroeconomic signals.

\begin{table}[b]
\centering
\caption{Agent-wise Signal Ablation on TSLA}
\label{tab:signal-ablation}
\small
\setlength{\tabcolsep}{2.5pt} % 全局收紧列间距
\begin{tabular}{@{} c c c c @{\quad} | c c c @{}}
\toprule
\multicolumn{4}{c}{\textbf{Enabled Signals}} 
& \multicolumn{3}{c}{\textbf{Performance}} \\
\cmidrule(lr){1-4} \cmidrule(lr){5-7}
Market & Company & 10-Q & 10-K 
& CR\% & SR& MDD\% \\
\midrule
\checkmark & \checkmark & \checkmark & \checkmark 
& \textbf{62.15} & \textbf{0.68} & \textbf{42.34} \\
\midrule
$\times$   & \checkmark & \checkmark & \checkmark 
& 51.32 & 0.53 & 55.40 \\
\checkmark & $\times$   & \checkmark & \checkmark 
& 48.15 & 0.48 & 60.52 \\
\checkmark & \checkmark & $\times$   & \checkmark 
& 58.59 & 0.63 & 45.11 \\
\checkmark & \checkmark & \checkmark & $\times$   
& 53.78 & 0.56 & 67.25 \\
\bottomrule
\end{tabular}
\end{table}

\subsubsection{Macroeconomic
News (MN) module}
\label{appendix:mn}
% 将加入宏观新闻后决策agent的思考过程可视化出来,然后说明宏观新闻为决策带来了哪些积极影响。
Beyond the experiments mentioned in the main text, FinPOS further broadens its information scope by incorporating macroeconomic news (MN) and other sources into the Market Signal Processing and Analysis module, and conducts corresponding ablation studies.
The following provides examples of the decision-making process of the directional agent before and after integrating the macroeconomic news module.

\begin{itemize}[left=0.3cm]
    \item \textbf{Before incorporating MN:}  
    I begin by reviewing latest earnings report and recent company news of TSLA. The quarterly results exceed market expectations, and Musk's announcement about expanding factory investment conveys a strong growth signal. Market sentiment around the company appears to be positive, and technical indicators show sustained upward momentum. Based on the alignment of these signals, I infer that TSLA's short-term trend is likely to be bullish. Decision: \textbf{buy}.

    \item \textbf{After incorporating MN:}  
    Recent news indicates that Trump has introduced new tariffs, inflation is accelerating, and export conditions for U.S. manufacturers are deteriorating. Although TSLA's fundamentals remain strong, these macro developments imply higher production costs and potential demand pressure. After integrating these broader signals, I reassess the outlook: systemic risk outweighs company-level optimism. Decision: \textbf{sell}.
\end{itemize}

This comparison demonstrates that MN provides crucial contextual signals for decision-making, especially during periods of high market volatility or systemic risk, significantly enhancing the agent's risk awareness and decision accuracy.

\subsection{FinPOS Performance under Extreme Market Conditions}

To evaluate the robustness of FinPOS under highly volatile market regimes, we conduct an additional stress test over the period from March to April 2025, during which market conditions exhibit pronounced instability. This interval coincides with a series of major macroeconomic and political events, including the U.S. presidential election and heightened geopolitical tensions, leading to frequent regime shifts and elevated tail risks.

Table~\ref{tab:extreme_market} reports the performance of different methods during this period. DRL-based approaches suffer substantial drawdowns and consistently negative risk-adjusted returns, indicating limited robustness to abrupt distributional shifts. Similarly, most LLM-based agents operating under discrete buy/sell/hold decision frameworks display unstable behavior, particularly during rapid price reversals and high-volatility episodes.
In contrast, FinPOS maintains strong and consistent performance across all evaluated stocks, achieving higher cumulative returns and Sharpe ratios while substantially reducing maximum drawdowns. Notably, the performance gap between FinPOS and baseline methods widens under extreme market conditions, where unmanaged position exposure can quickly amplify downside risk.
We attribute this robustness to FinPOS’s explicit modeling of continuous position evolution and risk-aware sizing.
By combining CVaR-based position control with multi-timescale reward feedback, FinPOS is able to gradually reduce exposure even when directional signals remain uncertain.
This capability is especially critical during black-swan-like events.

\begin{table*}[t]
\centering
\caption{Extended results on TSLA, AAPL, and COIN (Mar--Sep 2025).}
\label{tab:extreme_market}
\resizebox{\textwidth}{!}{%
\begin{tabular}{l
S S S
S S S
S S S}
\toprule
\multirow{2}{*}{\textbf{Models}}
& \multicolumn{3}{c}{\textbf{TSLA}}
& \multicolumn{3}{c}{\textbf{AAPL}}
& \multicolumn{3}{c}{\textbf{COIN}} \\
\cmidrule(lr){2-4} \cmidrule(lr){5-7} \cmidrule(lr){8-10}
& {CR\%$\uparrow$} & {SR$\uparrow$} & {MDD\%$\downarrow$}
& {CR\%$\uparrow$} & {SR$\uparrow$} & {MDD\%$\downarrow$}
& {CR\%$\uparrow$} & {SR$\uparrow$} & {MDD\%$\downarrow$} \\
\midrule
\textit{Baseline} & {} & {} & {} & {} & {} & {} & {} & {} & {} \\
Random
& -32.13 & -0.91 & 62.10
& -34.21 & -0.28 & 59.40
& -0.22  & -0.01 & 17.60 \\
\midrule
\textit{DRL} & {} & {} & {} & {} & {} & {} & {} & {} & {} \\
PPO~\cite{ppo}
& -73.76 & -0.56 & 69.88
& -58.33 & -0.55 & 61.40
& -2.04  & -0.04 & 17.80 \\
\midrule
\textit{LLM-Based Agents} & {} & {} & {} & {} & {} & {} & {} & {} & {} \\
FinGPT~\cite{FinGPT}
& -95.19 & -1.07 & 94.36
& -85.22 & -1.52 & 74.60
& -3.63  & -0.04 & 17.00 \\
FinAgent~\cite{finagent}
& -74.31 & -0.89 & 85.65
& -78.00 & -1.09 & 71.50
& -2.10  & -0.03 & 17.50 \\
FINMEM~\cite{finmem}
& -44.03 & -0.52 & 72.10
& -45.88 & -0.54 & 47.80
& 0.13   & 0.01  & 19.20 \\
FINCON~\cite{fincon}
& 7.76   & 0.38  & 59.13
& -16.02 & -0.13 & 29.03
& 7.35   & \bfseries 1.03 & 15.10 \\
\midrule
\textbf{FinPos}
& \bfseries 54.99 & \bfseries 0.67 & \bfseries 42.34
& \bfseries 60.28 & \bfseries 0.69 & \bfseries 19.75
& \bfseries 14.74 & 0.92 & \bfseries 14.05 \\
\bottomrule
\end{tabular}%
}
\end{table*}

%% file: custom.bib
@techreport{reinforcement,
  title={Reinforcement learning in financial markets-a survey},
  author={Fischer, Thomas G},
  year={2018},
  institution={FAU discussion papers in economics}
}

@article{DRL,
  title={Deep reinforcement learning for trading—A critical survey},
  author={Millea, Adrian},
  journal={Data},
  volume={6},
  number={11},
  pages={119},
  year={2021},
  publisher={MDPI}
}

@article{survey_DRL,
  title={A survey on deep reinforcement learning architectures, applications and emerging trends},
  author={Balhara, Surjeet and Gupta, Nishu and Alkhayyat, Ahmed and Bharti, Isha and Malik, Rami Q and Mahmood, Sarmad Nozad and Abedi, Firas},
  journal={IET Communications},
  year={2022},
  publisher={Wiley Online Library}
}

@article{news,
  title={How news affects the trading behaviour of different categories of investors in a financial market},
  author={Lillo, Fabrizio and Miccich{\`e}, Salvatore and Tumminello, Michele and Piilo, Jyrki and Mantegna, Rosario N},
  journal={Quantitative Finance},
  volume={15},
  number={2},
  pages={213--229},
  year={2015},
  publisher={Taylor \& Francis}
}

@article{FinGPT,
  author       = {Hongyang Yang and
                  Xiao{-}Yang Liu and
                  Christina Dan Wang},
  title        = {FinGPT: Open-Source Financial Large Language Models},
  journal      = {CoRR},
  volume       = {abs/2306.06031},
  year         = {2023},
  doi          = {10.48550/ARXIV.2306.06031},
  eprinttype    = {arXiv},
  eprint       = {2306.06031},
}

@inproceedings{finmem,
  title={FinMem: A performance-enhanced LLM trading agent with layered memory and character design},
  author={Yu, Yangyang and Li, Haohang and Chen, Zhi and Jiang, Yuechen and Li, Yang and Zhang, Denghui and Liu, Rong and Suchow, Jordan W and Khashanah, Khaldoun},
  booktitle={Proceedings of the AAAI Symposium Series},
  volume={3},
  number={1},
  pages={595--597},
  year={2024}
}

@inproceedings{finagent,
  title={A multimodal foundation agent for financial trading: Tool-augmented, diversified, and generalist},
  author={Zhang, Wentao and Zhao, Lingxuan and Xia, Haochong and Sun, Shuo and Sun, Jiaze and Qin, Molei and Li, Xinyi and Zhao, Yuqing and Zhao, Yilei and Cai, Xinyu and others},
  booktitle={Proceedings of the 30th ACM SIGKDD Conference on Knowledge Discovery and Data Mining},
  pages={4314--4325},
  year={2024}
}

@misc{agentmusic,
      title={MusicAgent: An AI Agent for Music Understanding and Generation with Large Language Models}, 
      author={Dingyao Yu and Kaitao Song and Peiling Lu and Tianyu He and Xu Tan and Wei Ye and Shikun Zhang and Jiang Bian},
      year={2023},
      eprint={2310.11954},
      archivePrefix={arXiv},
      primaryClass={cs.CL},
      url={https://arxiv.org/abs/2310.11954}, 
}

@article{medagent,
  title={Medagent-pro: Towards multi-modal evidence-based medical diagnosis via reasoning agentic workflow},
  author={Wang, Ziyue and Wu, Junde and Low, Chang Han and Jin, Yueming},
  journal={arXiv preprint arXiv:2503.18968},
  year={2025}
}

@misc{researchagent,
      title={Chain of Ideas: Revolutionizing Research Via Novel Idea Development with LLM Agents}, 
      author={Long Li and Weiwen Xu and Jiayan Guo and Ruochen Zhao and Xingxuan Li and Yuqian Yuan and Boqiang Zhang and Yuming Jiang and Yifei Xin and Ronghao Dang and Deli Zhao and Yu Rong and Tian Feng and Lidong Bing},
      year={2024},
      eprint={2410.13185},
      archivePrefix={arXiv},
      primaryClass={cs.AI},
      url={https://arxiv.org/abs/2410.13185}, 
}

@article{fincon,
  title={Fincon: A synthesized llm multi-agent system with conceptual verbal reinforcement for enhanced financial decision making},
  author={Yu, Yangyang and Yao, Zhiyuan and Li, Haohang and Deng, Zhiyang and Jiang, Yuechen and Cao, Yupeng and Chen, Zhi and Suchow, Jordan and Cui, Zhenyu and Liu, Rong and others},
  journal={Advances in Neural Information Processing Systems},
  volume={37},
  pages={137010--137045},
  year={2024}
}

@article{ta,
  title={TradingAgents: Multi-Agents LLM Financial Trading Framework},
  author={Xiao, Yijia and Sun, Edward and Luo, Di and Wang, Wei},
  journal={arXiv preprint arXiv:2412.20138},
  year={2024}
}

@article{CVaR,
  title={Optimization of conditional value-at-risk},
  author={Rockafellar, R Tyrrell and Uryasev, Stanislav and others},
  journal={Journal of risk},
  volume={2},
  pages={21--42},
  year={2000},
  publisher={Citeseer}
}

@article{gpt4O,
  title={Gpt-4o system card},
  author={Hurst, Aaron and Lerer, Adam and Goucher, Adam P and Perelman, Adam and Ramesh, Aditya and Clark, Aidan and Ostrow, AJ and Welihinda, Akila and Hayes, Alan and Radford, Alec and others},
  journal={arXiv preprint arXiv:2410.21276},
  year={2024}
}

@article{deepseek,
  title={Deepseek-v3 technical report},
  author={Liu, Aixin and Feng, Bei and Xue, Bing and Wang, Bingxuan and Wu, Bochao and Lu, Chengda and Zhao, Chenggang and Deng, Chengqi and Zhang, Chenyu and Ruan, Chong and others},
  journal={arXiv preprint arXiv:2412.19437},
  year={2024}
}

@article{RL,
  title={Reinforcement learning algorithms: A brief survey},
  author={Shakya, Ashish Kumar and Pillai, Gopinatha and Chakrabarty, Sohom},
  journal={Expert Systems with Applications},
  volume={231},
  pages={120495},
  year={2023},
  publisher={Elsevier}
}

@inproceedings{llm_decision,
  title={Enhancing decision-making for llm agents via step-level q-value models},
  author={Zhai, Yuanzhao and Yang, Tingkai and Xu, Kele and Feng, Dawei and Yang, Cheng and Ding, Bo and Wang, Huaimin},
  booktitle={Proceedings of the AAAI Conference on Artificial Intelligence},
  volume={39},
  number={25},
  pages={27161--27169},
  year={2025}
}

@inproceedings{ppo,
  title={Option-Driven Sentiment in FinRL: a PPO Approach to Trading},
  author={Li, Shenjian and Yu, Mingxuan and Dossor, Freddie},
  booktitle={2025 IEEE 11th International Conference on Intelligent Data and Security (IDS)},
  pages={62--64},
  year={2025},
  organization={IEEE Computer Society}
}

@article{dqn,
  title={Improving financial trading decisions using deep Q-learning: Predicting the number of shares, action strategies, and transfer learning},
  author={Jeong, Gyeeun and Kim, Ha Young},
  journal={Expert Systems with Applications},
  volume={117},
  pages={125--138},
  year={2019},
  publisher={Elsevier}
}

@inproceedings{a3c,
  title={An asynchronous advantage actor-critic reinforcement learning method for stock selection and portfolio management},
  author={Kang, Qinma and Zhou, Huizhuo and Kang, Yunfan},
  booktitle={Proceedings of the 2nd International Conference on Big Data Research},
  pages={141--145},
  year={2018}
}

@article{macd,
  title={Predicting stock price trend using MACD optimized by historical volatility},
  author={Wang, Jian and Kim, Junseok},
  journal={Mathematical Problems in Engineering},
  volume={2018},
  number={1},
  pages={9280590},
  year={2018},
  publisher={Wiley Online Library}
}

@article{rsi,
  title={Validity and reliability of the reflux symptom index (RSI)},
  author={Belafsky, Peter C and Postma, Gregory N and Koufman, James A},
  journal={Journal of voice},
  volume={16},
  number={2},
  pages={274--277},
  year={2002},
  publisher={Elsevier}
}

@article{finrl,
  title={FinRL: A deep reinforcement learning library for automated stock trading in quantitative finance},
  author={Liu, Xiao-Yang and Yang, Hongyang and Chen, Qian and Zhang, Runjia and Yang, Liuqing and Xiao, Bowen and Wang, Christina Dan},
  journal={arXiv preprint arXiv:2011.09607},
  year={2020}
}

@article{stock,
  title={Stock Trading Using a Deep Reinforcement Learning and Text Analysis},
  author={Benk, Dominik},
  year={2022},
  publisher={Univerzita Karlova, Fakulta soci{\'a}ln{\'\i}ch v{\v{e}}d}
}
